\documentclass[aps,prb,showpacs,twocolumn,notitlepage,superscriptaddress,nofootinbib]{revtex4-2}
\usepackage{stix}
\usepackage{bm,color,amsmath,graphicx,psfrag,accents,float}
\usepackage{multirow}
\usepackage{dcolumn}
\usepackage{xcolor}
\usepackage{comment}
\usepackage{tcolorbox}
\newcolumntype{L}[1]{>{\raggedright\arraybackslash}p{#1}}
\newcolumntype{C}[1]{>{\centering\arraybackslash}p{#1}}
\newcolumntype{R}[1]{>{\raggedleft\arraybackslash}p{#1}}

        \definecolor{AAcolor}{rgb}{0.7,0.1,0.4}

			\newcommand{\e}[1]{\begin{align}{#1}\end{align}}	
			\newcommand{\lin}{\notag \\}
		
		\newcommand{\f}[2]{\frac{#1}{#2}}
		\newcommand{\tf}[2]{\tfrac{#1}{#2}}
		
		\newcommand{\la}[1]{\label{#1}}

		\newcommand{\q}[1]{Eq.\ (\ref{#1})}
		\newcommand{\qq}[2]{Eqs.\ (\ref{#1})-(\ref{#2})}
		\newcommand{\s}[1]{Sec.\ \ref{#1}}
		\newcommand{\fig}[1]{Fig.\ \ref{#1}}		
		\newcommand{\app}[1]{App.\ \ref{#1}}				
		
		\newcommand{\ocite}[1]{Ref.\ \onlinecite{#1}}

		\newcommand{\ri}{\rightarrow}
		
		\newcommand{\lea}{\leftarrow}
				
		\newcommand{\sgn}{\text{sgn}}

		\newcommand{\eq}{=&\;}

		\newcommand{\Z}{\mathbb{Z}}

\newcommand{\mathsout}[1]
{\bgroup\mathchoice
  {\sbox0{$\displaystyle{#1}$}%
    \usebox0\hspace{-\wd0}%
    \rule[0.5\ht0-0.5\dp0-.5pt]{\wd0}{1pt}}%
  {\sbox0{$\textstyle{#1}$}%
    \usebox0\hspace{-\wd0}%
    \rule[0.5\ht0-0.5\dp0-.5pt]{\wd0}{1pt}}%
  {\sbox0{$\scriptstyle{#1}$}%
    \usebox0\hspace{-\wd0}%
    \rule[0.5\ht0-0.5\dp0-.5pt]{\wd0}{1pt}}%
  {\sbox0{$\scriptscriptstyle{#1}$}%
    \usebox0\hspace{-\wd0}%
    \rule[0.5\ht0-0.5\dp0-.5pt]{\wd0}{1pt}}%
\egroup}

	\newcommand{\eikr}{e^{i\bk \cdot \br}}

	\newcommand{\emikr}{e^{-i\bk \cdot \br}}

\newcommand{\nabk}{\nabla_{\boldsymbol{k}}}

\newcommand{\var}{\varepsilon}

\newcommand\as{\;\;\;\;}

\newcommand{\bd}{\boldsymbol{d}}

\newcommand{\bi}{\boldsymbol{i}}
\newcommand{\bj}{\boldsymbol{j}}
\newcommand{\bk}{\boldsymbol{k}}

\newcommand{\bn}{\boldsymbol{n}}
\newcommand{\bp}{\boldsymbol{p}}
\newcommand{\bq}{\boldsymbol{q}}
\newcommand{\br}{\boldsymbol{r}}

\newcommand{\bw}{\boldsymbol{w}}

\newcommand{\by}{\boldsymbol{y}}

\newcommand{\bA}{\boldsymbol{A}}

\newcommand{\bG}{\boldsymbol{G}}

\newcommand{\bR}{\boldsymbol{R}}
\newcommand{\bS}{\boldsymbol{S}}

\newcommand{\bze}{\boldsymbol{0}}

\newcommand{\beps}{\boldsymbol{\epsilon}}

\newcommand{\bOmega}{\boldsymbol{\Omega}}
\newcommand{\bsigma}{\boldsymbol{\sigma}}

\newcommand{\W}{{\cal W}}

\newcommand{\vectwo}[2]{\begin{pmatrix} {#1}\\{#2} \end{pmatrix}}

\newcommand{\sx}{\sigma_{\sma{1}}}
\newcommand{\sy}{\sigma_{\sma{2}}}
\newcommand{\sz}{\sigma_{\sma{3}}}

\newcommand{\cale}{{\cal E}}

\newcommand{\bcals}{\boldsymbol{\cal S}}

\newcommand{\scrs}{{\mathscr S}}

\newcommand{\noi}[1]{\noindent (#1)}

\newcommand{\braket}[2]{\big\langle #1 \,|\, #2 \big\rangle}

\newcommand{\braopket}[3]{\big\langle #1 \,|\, #2 \,|\, #3 \big\rangle}

\newcommand{\ket}[1]{|\,#1\,\rangle}

\newcommand{\ab}{\alpha\beta}

\newcommand{\bpm}{\begin{pmatrix}}
\newcommand{\epm}{\end{pmatrix}}

\newcommand{\bal}{\begin{align}}

\newcommand{\eps}{\epsilon}

\newcommand{\dg}[1]{#1^{\scriptstyle{\dagger}}}

\newcommand{\sma}[1]{\scriptscriptstyle{#1}}

\begin{document}

\title{Quantization of intra- and inter-band Berry phases in the shift current}

\author{A. Alexandradinata} \affiliation{Department of Physics and Santa Cruz Materials Center, University of California Santa Cruz, Santa Cruz, CA 95064, USA}


\begin{abstract}
The theory of the shift current is thus far geometrical without being topological. This means that the real-space displacement/shift of a photoexcited quasiparticle  depends on the geometric Berry phase, but the Berry phase is not quantized to a rational multiple of $2\pi$.  I rectify this status quo by introducing a new class of topological insulators whose band topology is \emph{only} compatible with a non-centrosymmetric space group.  For such insulators, it is impossible to continuously tune the $\bk$-dependent shift vector to zero throughout the Brillouin zone. Suitably averaged, the shift vector is quantized to a rational multiple of a Bravais lattice vector.  Even with wide band gaps, the frequency-integrated shift conductivity greatly exceeds $e^3/h^2$, and is at least three orders of magnitude larger than the conductivity of  the prototypical ferroelectric BaTiO$_3$.  The large conductivity is attributed to an interplay between quantized intra- and inter-band Berry phases. In particular, topological defects of the inter-band Berry phase can enhance the  shift current, even for unpolarized insulators with negligible intra-band Berry phase.   \end{abstract}
\date{\today}

\maketitle


\section{Motivation and results}

The uniform illumination of a homogeneous but non-centrosymmetric material generates a direct photocurrent.\cite{sturmanfridkin_book} Part of this photocurrent originates from the real-space displacement (or \textit{shift})  of photo\textit{excited} quasiparticles as they vertically transit between bands.\cite{belinicher_kinetictheory} A geometric theory of the  \textit{excitation shift} current has developed based on
geometric interpretations of the electron polarization\cite{zak_berryphase,kingsmith_polarization,resta1994} and the dipole matrix element\cite{ahn_riemanniangeometry}; the real-space shift has been related to a geometric Berry phase\cite{belinicher_kinetictheory,sipe_secondorderoptical,morimoto_nonlinearoptic,morimoto_excitonic} which  may take any generic value -- it is not symmetry-fixed to a rational multiple of $2\pi$. The present theory of the shift current is thus geometrical without being topological -- lacking the defining quality of quantization that is robust against perturbations.\footnote{Topological invariants exist for the circular photogalvanic effect\cite{dejuan_quantizedcircular} and the photovoltaic Hall effect\cite{ahn_riemanniangeometry}.} \\

\begin{figure}[h]
\centering
\includegraphics[width=7.5 cm]{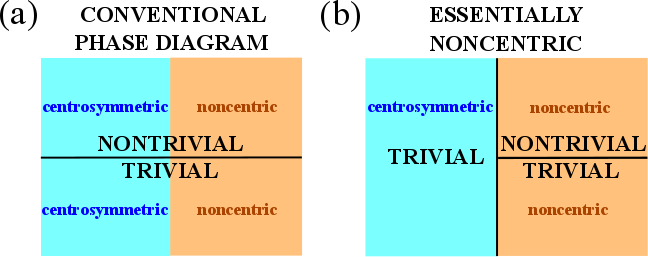}
\caption{(a) Phase diagram of conventional topological materials: the topologically nontrivial phase of matter straddles the boundary between centrosymmetric and non-centrosymmetric (i.e., `noncentric') Hamiltonians.  (b) Phase diagram of essentially noncentric topological materials: the topologically nontrivial phase of matter is only compatible with a noncentric Hamiltonian.}\label{fig:intrinsicallynoncentric}
\end{figure}

Why was no quantized geometric phase found in previous investigations\cite{ahn_lowfrequencydivergence,chingkit_photocurrentinweyl,xuyang_photovoltaicweyl,osterhoudt_colossalBVPEweyl,kim_shiftdiracsurface,LiangTan_topoins} of the excitation shift current in topological materials? Because it is  possible to continuously deform the insulating tight-binding Hamiltonian (or semimetallic low-energy Hamiltonian) to be centrosymmetric with vanishing shift current, while remaining in the same topological phase, as illustrated in \fig{fig:intrinsicallynoncentric}(a). This implies for the studied classes of topological materials  that  nontrivial topology of the wave function is not, by itself, a sufficient condition for a nontrivial shift; further supplemental conditions must be added to ensure the shift, e.g., proximity to a topological phase transition,\cite{LiangTan_topoins} or  tilting\cite{ahn_lowfrequencydivergence,chingkit_photocurrentinweyl,xuyang_photovoltaicweyl}/warping\cite{kim_shiftdiracsurface} of energy dispersions. \\

Aiming to forgo all supplemental conditions, this work introduces a new class of topological insulators for which wave-function topology is a sufficient condition for a nontrivial shift. The introduced class contrasts from previous case studies in being \emph{essentially noncentric}, meaning that the topologically nontrivial phase of matter exists \emph{only} in  crystal classes without a center of inversion, as illustrated in \fig{fig:intrinsicallynoncentric}(b). In other words, the lack of centrosymmetry is essential to meaningfully distinguish between phases that are topologically trivial vs nontrivial.  \\

The sufficient condition for a nontrivial shift reads as follows:\\

\noi{P1} For essentially noncentric topological insulators, a geometric quantity exists that inputs band wave functions and outputs an integer; if this integer is nonzero, the $\bk$-dependent photonic shift vector cannot be continuously tuned to zero throughout the Brillouin zone.\\

\noindent The \textit{photonic} shift vector $\bS^{\beps}_{b'\bk\lea b\bk}$  is  the real-space shift  of an electronic quasiparticle as it transits  from band $b$ to  band $b'$ (at fixed wavevector $\bk$), by way of emitting/absorbing   a \textit{photon} with linear polarization vector $\beps$.\footnote{The geometric interpretation of the phononic shift vector is discussed in a companion paper which focuses on the steady photovoltaic current.\cite{AAzhu_anomalousshift}}   In terms of the multi-band Berry connection $\bA_{b'b\bk}{=}\braket{u_{b'\bk}}{i\nabk u_{b\bk}}_{cell}$,\footnote{This inner product involves integrating the intracellular coordinate over the primitive unit cell, with the normalization  $\braket{u_{b\bk}}{u_{b'\bk}}_{cell}{=}\delta_{b,b'}$.} 
\e{\bS^{\beps}_{b'\bk \lea b\bk}\eq -\nabk \arg \beps \cdot \bA_{b'b\bk}+\bA_{b'b'\bk}-\bA_{bb\bk}.\la{defineshiftvector}}
We will refer to $\bA_{bb}$ as the \emph{intra-band Berry connection} for the $b$'th band; the \emph{inter-band Berry phase} $(\arg \beps \cdot \bA_{b'b\bk})$ is the phase/argument of the complex-valued, band-off-diagonal Berry connection, which enters the theory through the dipole-transition matrix element   $e\cale_{\omega} \beps{\cdot}\bA_{b'b}$, with $[\boldsymbol{\cale}(\br,t){=}\beps\cale_{\omega}e^{i(\bq\cdot\br-i\omega t)} +$complex conjugate]  being the incident electric wave. \\

\begin{figure}[h]
\centering
\includegraphics[width=8.6 cm]{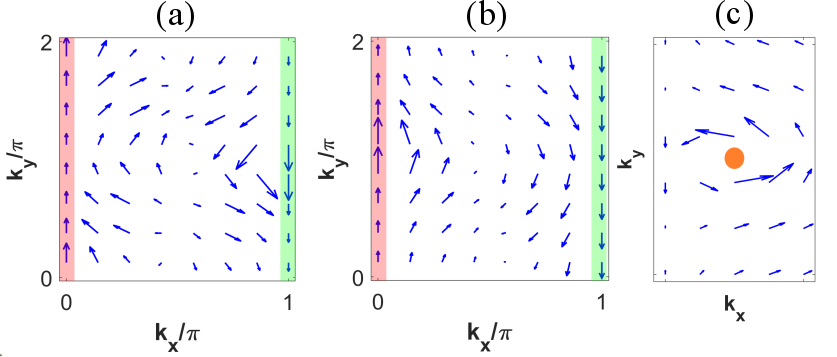}
\caption{Representative plots of the shift vector field for  a two-dimensional essentially noncentric insulator. The horizontal component of each arrow is proportional to $S^x_x$ and the vertical component to $S^x_y$.  The vector fields in panels  (a) and (b) are continuously deformable into each other and share identical topological invariants: averaging the shift vector over either reflection-invariant line (colored red and green) gives exactly a primitive Bravais lattice vector. (c) A vortex in the shift vector field indicates a topological defect of the inter-band Berry connection. }\label{fig:shiftvectorfield}
\end{figure}

If the shift vector is viewed as a vector field over $\bk$-space, Proposition (P1) implies there exists topologically nontrivial fields which cannot be continuously deformed to the zero vector field; two representative examples are illustrated in \fig{fig:shiftvectorfield}(a) and (b). The `geometric quantity' in Proposition (P1) is expressed in \q{shiftobstruction} as a sum of a quantized intra-band Berry phase and a quantized inter-band Berry phase; the latter quantity is associated to topological defects (in momentum space) of the inter-band Berry connection, as illustrated in \fig{fig:shiftvectorfield}(c). \\

In evocative terms, the topological knot of the electronic wave function carries an unremovable polarity; in precise terms:\\

\noi{P2} For an essentially noncentric insulator with a reflection symmetry,  averaging the shift vector over either reflection-invariant $\bk$-plane gives \emph{exactly} a Bravais lattice vector. \\

\noindent There being two such $\bk$-plane gives two independent vectors: $\bcals^{\text{ave}}_{0}$ and $\bcals^{\text{ave}}_{\pi/R_x}$.  
The direction of $\Sigma\bcals_{\text{ave}}{:}{=}\bcals^{\text{ave}}_{0}{+}\bcals^{\text{ave}}_{\pi/R_x}$ may be interpreted as the  polar axis of the electronic wave function. A nonzero $\Sigma\bcals_{\text{ave}}$
connects different primitive unit cells and may be described as \emph{intercellular}.  The associated shift current is expected to be larger than   in existing shift-current materials where  intracellular charge transfer occurs between atoms in one unit cell.\cite{nastos_opticalrectification,braun_ultrafastphotocurrents} The analogs of $\bcals^{\text{ave}}_{0}$ and $\bcals^{\text{ave}}_{\pi/R_x}$ for two-dimensional insulators are obtained by averaging the shift vector over reflection-invariant $\bk$-lines, as illustrated in \fig{fig:shiftvectorfield}(a-b).  \\

Propositions (P1) and (P2) are topological principles to guide the search of materials with large shift currents. 
To quantify how large, I will use a figure of merit  expressed in terms of the fundamental geometric quantity  -- the \emph{photonic shift connection}:\cite{ahn_riemanniangeometry}
\e{C^j_{ib'b\bk}=|A_{jb'b\bk}|^2 S^{j}_{ib'b\bk};\as S^j_{ib'b\bk}=\vec{\bi}\cdot\bS^{\vec{\bj}}_{b'\bk\lea b\bk}.} 
In our adopted shorthand for the shift vector, $i,j \in \{x,y,z\}$ label the Cartesian axes and $\vec{\bi},\vec{\bj}$ are unimodular directional vectors. Likewise, $A_{jb'b\bk}{=}\vec{\bj}{\cdot}\bA_{b'b\bk}.$
The shift connection enters the expression of the excitation shift current  as the average velocity (in the $i$'th direction) of a shifting quasiparticle:
$e^3|\cale_{\omega}^2|C_{icv}^{j}\delta(\hbar\omega{-}\var_c{+}\var_v)/\hbar,$
given by the shift vector multiplied by the photoexcitation transition rate, namely the rate an electron is photoexcited from a fully-filled valence band (with energy $\var_v$) to a fully-empty conduction band $(\var_c)$ by a monochromatic light source with photon energy $\hbar\omega$ and linear polarization $\vec{\bj}$. Motivated by a broadband light source (e.g., solar light) with a spectral peak that is as wide as a typical band, I integrate the rate over all optical excitations between  the two bands lying closest to the Fermi level; the resultant quantity is proportional to the Brillouin-zone-integral of the shift connection:
\e{ F_{i}^j=\int_{BZ} d^3k C^{j}_{icv}.  \la{bandshift}}
I adopt $F_{i}^j$ as a dimensionless figure of merit. This figure is proportional to the frequency-integrated \textit{excitation shift conductivity}:\footnote{The prefactor of $2$ in \q{zeroT} reflects the spin degeneracy of bands in insulators with negligible spin-orbit coupling. $\sigma^{\text{exc},j}_{i\omega}$ translates to $\vec{\bi}\cdot \bsigma^{\text{exc}}_{\vec{\bj}\omega}$ in the companion paper.\cite{AAzhu_anomalousshift}} 
\e{ 2 F_i^j \f{e^3}{h^2} = \int \sigma^{\text{exc},j}_{i\omega}\bigg|_{T=0}d\omega. \la{zeroT}}
This nonlinear conductivity is defined through the excitation shift current: $j_i^{\text{exc},j}=\sigma^{\text{exc},j}_{i\omega}|\cale_{\omega}|^2$; the subscript $({T=0})$ reminds us that we are photoexciting a zero-temperature insulator. The excitation shift conductivity is closely related to a measurable transient photocurrent, as elaborated in \s{sec:discussion}. \\

For essentially noncentric insulators with a band gap $E_g$ (minimized over the BZ), a band width $E_w$ (maximized between conduction and valence band), a polar axis parallel to $y$, and a  reflection symmetry mapping $x{\ri}{-}x$, I propose that:\\

\noi{Q1} For $E_g \gtrsim E_w$, $|F_y^x|\gg 1$ and is roughly proportional to the magnitude of the intercellular shift vector $\Sigma\bcals_{\text{ave}}$.\\ 

\noindent  Because  $\Sigma\bcals_{\text{ave}}$ is equivalently viewed as a $\Z^2$-valued invariant taking values in a two-dimensional (2D) Bravais lattice with all lattice periods set to unity, proposition (Q1)   epitomizes a maxim that to maximize the excitation shift current is to maximize a topological invariant. (Q1) also challenges a widely-held expectation that small band gaps are necessary for large excitation shift currents in topological materials.\cite{ahn_lowfrequencydivergence,chingkit_photocurrentinweyl,xuyang_photovoltaicweyl,LiangTan_topoins,LiangTan_upperlimit}   Because being topologically nontrivial is a global property of the entire band, the largeness of $\sigma_{y\omega}^{\text{exc},x}$ extends over a frequency range that is potentially comparable to the band width; this makes wide-gap essentially noncentric insulators suited for photoexcitation by solar light, since the solar spectrum has a broad peak covering ${2}$ to 3 eV.\footnote{The potential for shift-current materials as solar cells is discussed in a companion paper.\cite{AAzhu_anomalousshift}}\\

\noi{Q2} For $E_g \ll E_w$, $F_y^x$ diverges as $|E_g|^{-1/2}$ in the approach to a topological phase transition.  \\

\noindent This suggests an application to ultrafast infrared detection without an external bias voltage, which obviates the problem of the dark current in semimetallic photodetectors.\cite{liu_semimetalphotodetector}\\

 To  compare with known/predicted values for $F^{j}_i$, Tan and Rappe have computed (by first principles) the longitudinal $F^{j}_i$ for 950 noncentrosymmetric, nonmagnetic materials,\cite{LiangTan_upperlimit} finding: (a)  $|F^{y}_y|{\approx} 10^{-2}$ for the prototypical ferroelectric insulator $BaTiO_3$, with  $y$ parallel to the polar axis, and (b) $|F^{y}_y|{\approx} 3$ for SrAlSiH represents the best-performing insulator with $E_g{\geq}1 eV$; the former material has been experimentally benchmarked,\cite{youngrappe_firstprinciples,koch_BaTiO3} but not the latter.\\

For a further comparison with typical values of $\sigma_{i\omega}^{\text{exc},j}$, let us assume that an essentially noncentric insulator has a  band width of $E_w{=}1 eV$ and that $\Sigma\bcals_{\text{ave}}$ is proportional to a primitive Bravais lattice vector $\vec{B}$, with $||\vec{B}||{=}R_y$. Proposition (Q1) then  implies that the magnitude of the \emph{frequency-averaged} shift conductivity 
\e{|\langle \sigma_{y\omega}^{\text{exc},x} \rangle_{\text{ave}}|\gtrsim 0.1 mAV^{-2} \times\f{||\Sigma\bcals_{\text{ave}}||}{R_y},} with $||\Sigma\bcals_{\text{ave}}||/R_y$ an integer-valued topological multiplier.  In contrast, the largest \emph{peak} value $\sigma_{i\omega}^{\text{exc},j}$ among five polar compounds $\{X$TiO$_3$($X{=}$Ba,Pb), LiAsS$_2$, $Y$AsSe$_2$($Y{=}$Li,Na)$\}$ was calculated to be $0.05 mAV^{-2}$ in magnitude.\cite{youngrappe_firstprinciples,brehm_LiAsS2}\\

The comparative largeness of $|F_y^x|$ (for essentially noncentric insulators) originates from an interplay between the intra- and inter-band Berry phases: 
a large intra-band Berry phase does not necessarily result in a large shift current if topological defects of the inter-band Berry phase are present; conversely, a large shift current can be solely attributed to these topological defects --  for insulators with trivial intra-band Berry phase. Such interplay has not been considered in previous works\cite{fregoso_opticalzero,Schankler_MoS2} which maximize the shift current solely by optimizing the polarization, which is closely related to the intra-band Berry phase.\cite{zak_berryphase,kingsmith_polarization,resta1994} Only with a unified characterization of both intra- and inter-band Berry phases can one achieve a complete topological theory of the shift current.\\

Such a theory is developed in \s{sec:theory}, with the goal of establishing 
propositions (P1-2) for essentially noncentric insulators. \s{sec:models} presents two model Hamiltonians of essentially noncentric insulators to corroborate propositions (Q1-2). The theory and models will first be established in the simplest possible context: a point group generated by a single reflection, a Bravais lattice with a monatomic basis, and a low-energy Hilbert space given by two bands.
  The last Section [\s{sec:discussion}] recapitulates our results with a different set of motivations, as well as elaborates on experimental implications for the transient and steady photovoltaic currents. I end the paper by suggesting guidelines for an ab-initio-based, high-throughput search for noncentric insulators with nontrivial optical vorticity, and a different set of guidelines to search for essentially noncentric insulators. \\

An Appendix clarifies some mathematical niceties as well as generalizes the theory and models in the main text. \app{app:invariants} presents a rigorous formulation of a topological invariant that depends not only on the intra-band Berry connection, but also on the inter-band Berry connection. \app{sec:extensions} extends the theory beyond the simplifying assumptions made in the main text; in particular, the extension to $(N{>}2)$ bands leads naturally to identifying essentially noncentric insulators as having `delicate topology'.\cite{nelsonAA_multicellularity,nelsonAA_delicatetopology}
	Throughout this work, I employ the tight-binding approximation for the Berry and shift connections, which is generally an uncontrolled approximation; \app{sec:TBapprox} discusses how the approximation may be justified, as well as highlights an under-appreciated pitfall.

\section{Theory of essentially noncentric insulators}\la{sec:theory}

Let us attempt to deduce the geometrical invariants of an essentially noncentric insulator from basic principles.  One clue to determining the geometric quantity alluded to in proposition (P1) is that a nontrivial shift requires\cite{sturmanfridkin_book} the absence of spatial centrosymmetry. Let us therefore imagine what the geometry of band wave functions would look like, if these wave functions were to  \textit{maximally} break centrosymmetry, in a manner of speaking. More precisely, by viewing the Berry curvature $(\bOmega_{v\bk}{=}\nabla \times \bA_{vv\bk})$ and the band-off-diagonal Berry connection as geometrical vector fields over momentum space, we will try to concoct fields that do the opposite of what centrosymmetry imposes.

\subsection{Berry-curvature invariant that breaks centrosymmetry}

Because the curvature tranforms  as a pseudovector under crystallographic point-group operations, $\vec{z}{\cdot}\bOmega_{v\bk}{=} \Omega_{zv(k_x,k_y)}{=}{+}\Omega_{zv(-k_x,-k_y)}$ holds for any two-dimensional, centrosymmetric insulator; the theory will be extended to three dimensions later. To `maximally'  break centrosymmetry, let me (i) invert the sign in the centrosymmetry constraint to obtain:   $\Omega_{zv\bk}{=}{-}\Omega_{zv,-\bk}$, and (ii) ask that  the curvature integral [over \emph{half} the Brillouin zone (BZ)] be quantized to a nontrivial integer:
\e{ RTP_v:=\int_{BZ/2}\Omega_{zv} \f{d^2k}{2\pi} \in \Z.  \la{defineRTP}}

The first condition is guaranteed by time-reversal symmetry; the sign difference in the symmetry constraints originates from time reversal having an antiunitary\cite{Wigner_ontheoperationoftimereversal} representation $\hat{T}$ squaring to the identity, in contrast with the unitary representation of spatial inversion. \\ 

The second condition [\q{defineRTP}] is possible if one introduces a reflection symmetry: $x {\ri}{-}x$ and specifies $BZ/2$ to be the positive-$k_x$ half of the BZ. (The integral of the curvature over the negative-$k_x$ half of the BZ simply equals minus $RTP_v$ due to time-reversal symmetry.) To specify the action of reflection symmetry, I consider a reduced Hilbert space given by the highest-energy valence band and the lowest-energy conduction band, and assume that this Hilbert space is spanned by two basis Wannier orbitals per primitive unit cell.
(The restriction to two bands simplifies the initial presentation, but will be relaxed in \app{sec:extensions}.) Picking one representative unit cell, the two Wannier orbitals are labelled $\varphi_e$ and $\varphi_o$, with the subscript indicating that one orbital is reflection-even and the other reflection-odd; I assume for now  that both $\varphi$ are centered at the same location, such that all the `Wannier centers' form a rectangular lattice with a single-site basis and with periods $R_x$ and $R_y$ in the $x$ and $y$ directions respectively -- this being a natural assumption if the two Wannier orbitals are atomic orbitals of the same atom. (The assumption of a single-site basis  will also be relaxed in \app{sec:extensions}.) These assumptions on the Wannier orbitals translate to a symmetry constraint   $\sz h(\bk)\sz {=}h(-k_x,k_y)$ on the $\bk$-periodic, two-by-two matrix  Hamiltonian  $h(\bk)$, with $\sigma_3$ the Pauli matrix representation of reflection.  Eigenstates of $h(\bk)$ are denoted $\ket{u_{b\bk}}$ with corresponding energies $\var_{b\bk}$, with $b{=}v$ (resp.\ $c$) for the valence (resp.\ conduction) band, and $\var_c{>}\var_v$ for all $\bk$. \\

%

\begin{figure}[H]
\centering
\includegraphics[width=8.6 cm]{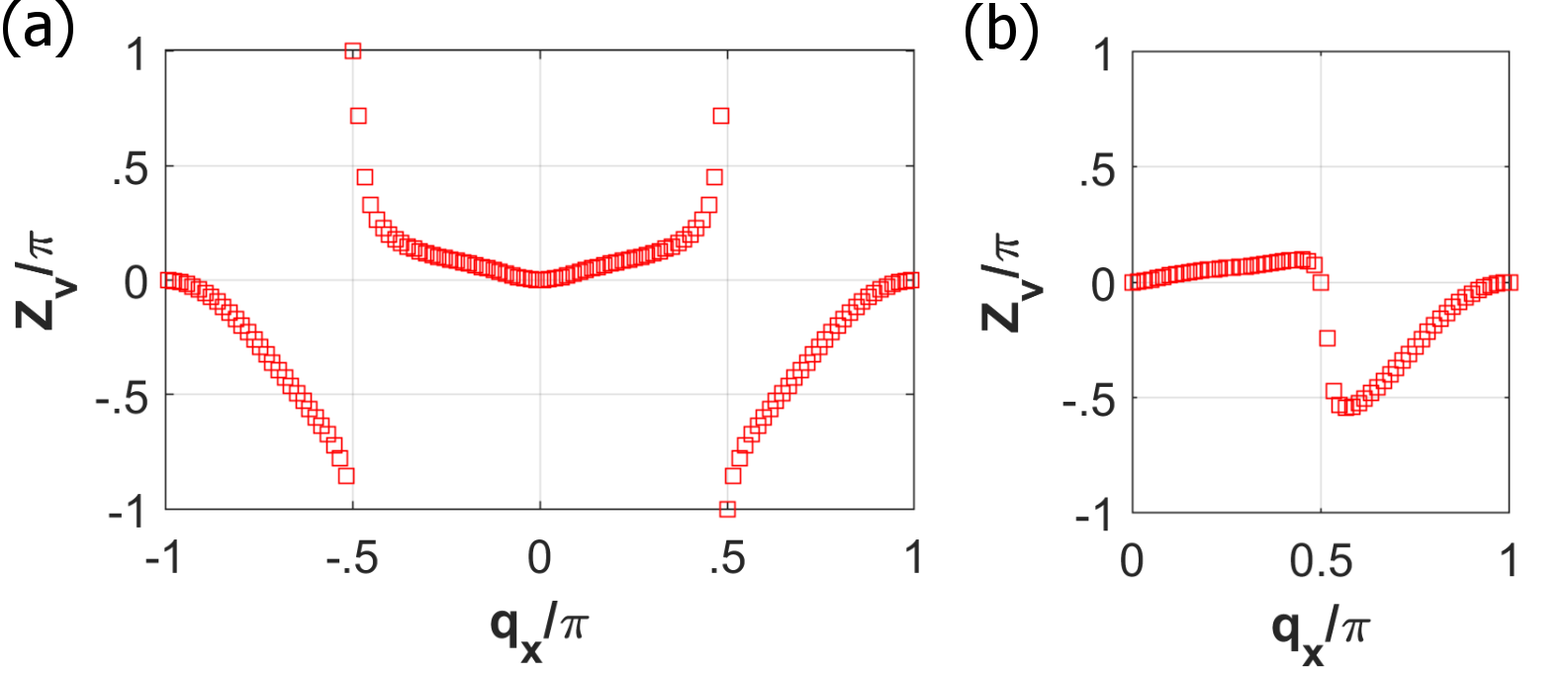}
\caption{(a) A reverting Thouless pump is revealed by a nontrivial dispersion of the valence-band Zak phase $Z_{v,k_x}$ [\q{definezakphase}]; $Z_v$ was computed with parameters $\alpha{=}2\beta{=}0.9$ in the model of \s{sec:modelsecondclass}. (b) A trivial pump for $\alpha{=}2\beta{=}1.1$.}\label{fig:zakphase}
\end{figure}

\noindent \emph{Proof that $RTP_v$ is integer-valued:} Stoke's theorem allows to equate    $RTP_v{=}[Z_{v,\pi/R_x}-Z_{v,0}]/2\pi$, with $Z_{b,k_x}$ the Berry-Zak phase acquired by parallel-transporting a Bloch state in band $b\in \{v,c\}$ over a $\bk$-loop with fixed $k_x$:
\e{ Z_{b,k_x}=\oint A_{ybb(k_x,k_y)} dk_y, \la{definezakphase}}
with $A_{bb\bk}$ the intra-band Berry connection for the tight-binding eigenstate $\ket{u_{b\bk}}$.   Denoting the parity (even vs odd) of a mirror-invariant Bloch state in band $b$ by $p(b,k_x)$, it follows from a known relation\cite{zak_berryphase} between the Berry-Zak phase  and the positional center of Wannier orbitals that
\e{ \text{for}\; k_x=0 \;\text{and} \;\frac{\pi}{R_x},\as \f{Z_{b,k_x}}{2\pi} =_1 \f{y[\varphi_{p(b,k_x)}]}{R_y},\la{zakwannier}} 
with $=_1$ denoting an equality modulo one, and $y[\varphi_e]$ the y-positional center of  the reflection-even basis Wannier orbital. The  assumption of a single-site basis guarantees that $y[\varphi_{p(b,0)}]=_{R_y} y[\varphi_{p(b,\pi/R_x)}]$, implying that $Z_{b,\pi/R_x}{-}Z_{b,0}$  can only be an integer multiple of $2\pi$, with this integer uniquely defined by insisting that the wave function is analytic over $BZ/2$. A representative, nontrivial example of the Berry-Zak phase is plotted in \fig{fig:zakphase}(a), with $Z_{b,k_x}$ continuously increasing by $2\pi$ as $k_x$ is advanced from $0$ to $\pi/R_x$; the reflection symmetry guarantees\cite{Cohomological} that $Z_b$ reverts to its original value upon further advancing $k_x$ by $\pi/R_x$. Viewing $k_x$ as an adiabatic parameter, $Z_{b,k_x}$ represents the pumping of one quantum of charge over half an adiabatic cycle, and a reverse pump over the next half.  This may be called a \emph{reverting Thouless pump},\footnote{Reverting pumps have been previously studied in contexts unrelated to nonlinear optics.\cite{AA_teleportation,nelsonAA_multicellularity,nelsonAA_delicatetopology}} in contrast with the non-reverting pumps studied by Thouless.\cite{thouless_pump} \\  

\subsection{Optical vortices break centrosymmetry}

For the inter-band Berry connection field $A_{xcv\bk}$,  centrosymmetry is `maximally' broken by introducing vortices, namely, quantized circulations of the phase  field $\arg A_{xcv}$ around a $\bk$-point where optical transitions vanish. For clarification, consider that $|A_{xcv}|^2$  is proportional to the probability transition rate  of resonant light absorption; I  refer to $|A_{xcv}|^2$ as the \emph{optical affinity} between conduction and valence bands; unlike the dipole-transition matrix element or the inter-band connection $A_{xcv}$, the affinity $|A_{xcv}|^2$ is gauge-invariant, i.e., unchanging under transformation of $\ket{u_{b\bk}}$ by a $\bk$-dependent phase factor. $\bk$-points where the affinity vanishes are called \emph{optical zeros}. 
\emph{Optical vortices} are optical zeros surrounded by a nontrivially circulating phase field.\footnote{The first example of an optical zero that is not an optical vortex is discussed in \s{sec:opticalphasetransition}. Though the phase  of $A_{xcv}$ is gauge-dependent, the circulation of the phase around a vortex is gauge-invariant, assuming that the gauge transformation $\ket{u_{b\bk}}{\ri}\ket{u_{b\bk}}e^{i\theta_{b\bk}}$ preserves the analyticity of $\ket{u_{b\bk}}$ with respect to $\bk$. The only way to change the phase circulation is with a discontinuous gauge transformation. Note that for insulators with trivial Chern invariants, the existence of wave functions which are analytic (with respect to $\bk$) and periodic over the Brillouin torus is guaranteed by the Grauert-Oka theorem; see references in footnote 12 of \ocite{nogo_AAJH}.}  \\

To visualize the circulation of the phase field, it is useful to introduce a Hamiltonian-vector interpretation of optical zeros and vortices: without loss of generality, I express $h(\bk){=}\bd(\bk){\cdot} \bsigma {+}h_{id}(\bk)I_{2\times 2}$ as a dot product of a real three-vector $\bd$ (the \emph{Hamiltonian vector}) with $\bsigma{:}{=}(\sx,\sy,\sz)$, plus a term proportional to the two-by-two identity matrix. Applying the identity
\e{A_{xcv}{=}\braopket{u_c}{\partial_{k_x}h}{u_v}_{cell}/i(\var_c-\var_v), \la{Acvidentity}} 
one deduces that an optical zero (with a nonzero energy gap) exists if and only if $\bd{\times}\partial_{k_x}\bd{=}0$. Since two real parameters (two spherical angles) need be tuned to align a vector $\bd$ to be collinear with $\partial_{k_x}\bd$, optical zeros 
generically form $(d{-}2)$-dimensional submanifolds of the $d$-dimensional BZ. For $d{=}2$, let us suppose an optical zero exists at the isolated wavevector $\bk_0$. For $\bk$ slightly deviating from $\bk_0$, $\bd$ and $\partial_{k_x}\bd$  slightly deviate from being collinear. If $\bk$ is advanced in a small circle around $\bk_0$, the two vectors maintain their non-collinearity and are able to rotate relative to each other, like two partner dancers locked in the closed position.\footnote{https://en.wikipedia.org/wiki/Closed$\_$position} The relative rotation of  $\partial_{k_x}\bd$ around $\bd$ (as $\bk$ makes a full circle) defines an integer-valued rotation number that is equivalent to the winding number  of the phase field $\arg A_{xcv}$. \\

Because of the unitary-antiunitary distinction in the representations of spatial and temporal inversions, the former symmetry constrains $A_{xcv\bk}{\propto}A_{xcv,-\bk}$ (with a proportionality phase factor that is analytic in $\bk$), while the latter symmetry constrains $A_{xcv\bk}{\propto}\overline{A_{xcv,-\bk}}$ (with the accent denoting complex conjugation). It follows  that centrosymmetry-related vortices have the same circulation while time-reversal-related vortices have the opposite. Thus, the presence of any optical vortex in a time-reversal-invariant Hamiltonian implies that centrosymmetry is broken.    

\subsection{Shift obstruction relation}

Having identified two topological quantities that are fundamentally incompatible with centrosymmetry, I now relate their linear combination to an integral of the shift vector:
\e{ Vort_x +2 {RTP}_v= -\Delta \scrs=\scrs_{0}-\scrs\big(\pi/R_x\big),\la{shiftobstruction}}
with $Vort_x$ (the \emph{net optical vorticity}) defined as the net circulation of all vortices of $A_{xcv}$ in $BZ/2$,\footnote{The net vorticity is uniquely defined by  
\e{ Vort_x=\int \partial_{k_y}\arg A_{xcv(\pi/R_x,k_y)}\tf{dk_y}{2\pi}-\int \partial_{k_y}\arg A_{xcv(0,k_y)}\tf{dk_y}{2\pi},} with conduction-band and valence-band wave functions that are analytic over $BZ/2$. If one allows for the wave function to be defined over  patches that cover $BZ/2$ and are mutually related by transition functions,\cite{bernevig_book}  the net vorticity loses its unique definition. }
and  $\scrs(k_x)$ defined as the \emph{line-averaged shift} (in units of the lattice period $b$) of all quasiparticles with wavenumber $k_x$,
\e{ \scrs(k_x)\eq \oint S_{ycv(k_x,k_y)}^x\f{dk_y}{2\pi}\lin
 \eq \f{Z_{c,k_x}-Z_{v,k_x}}{2\pi}-\oint \partial_{k_y}\arg A_{xcv} \f{dk_y}{2\pi}. \la{averageshift}}
$\Delta \scrs$ is thus the difference in line-averaged shifts between the two  mirror-invariant $\bk$-lines.\footnote{One can equivalently view $\Delta \scrs$ as the line integral (or circulation) of the shift vector along a $\bk$-rectangle whose two (of four) sides are mirror-invariant. Quantized circulations of an analogous shift vector have previously been studied in the context of interfacial reflection.\cite{liuying_quantizedcirculationshift} } In deriving \q{shiftobstruction}, I applied  that time-reversal symmetry guarantees the existence of Bloch functions (for both bands) that are 
analytic and periodic functions of $\bk$,\cite{Panati_trivialityblochbundle} hence
  $A_{xcv}$ is a meromorphic function of $\bk$ with discontinuities only at the optical vortices, and assumed in a generic situation that no vortices lie at a mirror-invariant wavevector;  use was also made of the complementary relation between the curvatures of conduction and valence bands: $\Omega_{zv\bk}{=}{-}\Omega_{zc\bk}$,\cite{spinorbitfree_AAchen} which leads to $RTP_v{=}{-}RTP_c$. Let me further remark on \qq{shiftobstruction}{averageshift}:\\

\noi{a} A closer inspection of \q{averageshift} reveals that  the line-averaged shift is integer-valued for mirror-invariant values of $k_x$. This follows from substitution of \q{zakwannier}, with $y[\varphi_e]{=}y[\varphi_o]$ guaranteed by the lattice basis being monatomic. Combining this result with the symmetry constraint that $S^x_{xcv\bk}$ vanishes at all mirror-invariant wavevectors, we deduce that optically-excited quasiparticles with $k_x{=}0$ are shifted by exactly $\scrs_{0}$ primitive lattice vectors parallel to the polar axis, on average. We thus arrive at proposition (P2), with the intercellular shift vectors: $\bcals^{\text{ave}}_{0}{=}\scrs_{0}R_y\vec{\by}$ and $\bcals^{\text{ave}}_{\pi/R_x}{=}\scrs_{\pi/R_x}R_y\vec{\by}$. (I will refer to the dimensionless scalars $\scrs_{0}$ and $\scrs_{\pi/R_x}$ as `intercellular shifts', and $\Delta \scrs$ as the `relative intercellular shift'.) It is worth emphasizing that the averaging process is essential for quantization, i.e., the shift vector at any specific $\bk$ is not quantized.\\

\noi{b}  Suppose $Vort_x{+}2 {RTP}_v$ in \q{shiftobstruction} is nonzero, and one has the ability to perturb the tight-binding Hamiltonian $h(\bk)$ and therefore modify $S_{ycv\bk}^x$. Despite  $S_{ycv\bk}^x$ being modifiable at each $\bk$, there exists a continuous range of possible perturbations where $Vort_x{+}2 {RTP}_v$ is invariant.\footnote{The conditions that preclude a discontinuous change in  $Vort_x{+}2 {RTP}_v$ are discussed in \app{app:invariants}.} One would then encounter a \emph{shift obstruction}: a topological obstruction against continuously tuning the shift vector to zero for all $\bk$; this is proposition (P1) in the introduction. For this reason I refer to \q{shiftobstruction} as the \emph{shift obstruction relation}.\footnote{While all quantities in the shift obstruction relation [\q{shiftobstruction}] were derived to be integer-valued for essentially noncentric insulators, actually \q{shiftobstruction} holds for any two-band insulator -- with the caveat that $RTP_v$ and $\Delta\scrs$ generically deviate from integer values, thus precluding a shift obstruction.}\\

\noi{c} A 2D reflection-symmetric insulator with $RTP_v{=}0$ is deemed topologically trivial under every known classification scheme based on the intra-band Berry connection: stable topology,\cite{kitaev_periodictable,shiozaki_review,bandcombinatorics_kruthoff} fragile topology,\cite{po_fragile,zhida_fragileaffinemonoid,bouhon_wilsonloopapproach,crystalsplit_AAJHWCLL} delicate topology,\cite{nelsonAA_multicellularity,nelsonAA_delicatetopology} topological quantum chemistry,\cite{TQC} symmetry-based indicators,\cite{Po_symmetryindicators} and wilson-loop characterizations.\cite{AA_wilsonloopinversion,crystalsplit_AAJHWCLL,bouhon_wilsonloopapproach,bradlyn_disconnectedEBR} What the shift obstruction relation reveals is that even such `trivial' insulators can have a nontrivial inter-band optical vorticity, implying that at least one of the two intercellular shifts is nonzero. Conversely, being topologically nontrivial (in the common usage of these words) is not a sufficient condition for a shift obstruction, because it is possible for the inter-band-Berry-phase contribution ($Vort_x$) to cancel out the intra-band-Berry-phase contribution ($RTP_v$).

\subsection{Implications for the photonic shift connection}\la{sec:implications}

What directly enters expressions for the excitation shift current is  the photonic shift connection  $C^{x}_{ycv}{=}|A_{xcv}|^2S_{ycv}^x$, whose value I now estimate for essentially noncentric insulators. An estimate is also presented for our figure of merit: the BZ-integrated shift connection [\q{bandshift}].\\

\noi{i} $E_g {\gtrsim} E_w$: If one is not close to a band-gap-closing, topological phase transition, the characteristic scale of variation for the optical affinity is the  BZ period. I therefore estimate the BZ-averaged optical affinity as  $\langle|A_{xcv}|^2\rangle{\sim} (R_x/2\pi)^2$ by dimensional analysis, with $(R_x,R_y,R_z)$ being the lattice period in the $(x,y,z)$ directions respectively. (This estimate is not affected by the possible existence of optical zeros, which occupy a measure-zero subregion of the BZ.) Assuming the \emph{average intercellular shift} $\langle\scrs\rangle{=}[\scrs_{0}{+}\scrs_{\pi/R_x}]/2$ is nonzero and independent of $k_z$, the BZ-averaged shift vector is estimated as $\langle S^x_y\rangle {\sim}\langle\scrs\rangle R_y$.  Then our figure of merit [cf. \q{bandshift}]   $F_y^x {\sim} \int d^3k \langle|A_{xcv}|^2\rangle \langle S^x_y\rangle{=}2\pi \langle\scrs\rangle R_x/R_z$. This is a plausibility argument to support proposition (Q1), with the identification $\Sigma \bcals_{\text{ave}}{=}2\langle\scrs\rangle R_y\vec{\by}$. If $\langle\scrs\rangle$  were to vanish but not the individual intercellular shifts, then $C^{x}_{ycv\bk}{\sim}R_x^2 R_y\scrs_{0}/(2\pi)^2$ for $k_x{\approx}0$. These estimates will be corroborated by model Hamiltonians in the next Section. \\

\noi{ii} $E_g {\ll} E_w$: Close to a topological phase transition, the minimal band gap $(E_g)$ over the BZ enters as a new scale in the problem. Most directly, it enters in the denominator of \q{Acvidentity}, leading to a divergence of the optical affinity for $\bk$ at the band-touching point; less directly, $\braopket{u_c}{\partial_{k_x}h}{u_v}_{cell}$ in the numerator of \q{Acvidentity} may also depend implicitly on $E_g$. The net effect of the explicit and implicit dependences is that the optical affinity may diverge as $|E_g|^{-2+\alpha}$, with $\alpha{\geq}0$.  I distinguish between \emph{first-class} phase transitions where the optical affinity diverges as $|E_g|^{-1}$ and \emph{second-class} transitions where the affinity  diverges as $E_g^{-2}$. Due to these divergences, $F_y^x$ may potentially also diverge and greatly exceed the estimates made for $E_g {\gtrsim} E_w$ in the previous paragraph; but this is not  self-evident a priori, because of the potentially-nontrivial $\bk$-dependence of the shift connection near the wavevector of closest inter-band contact. Two models will be presented in the next Section: one for which the phase transition is second-class but $F_y^x$ does not diverge (and instead displays a weaker kink-type non-analyticity), and a second model for which the phase transition is first class and   $F_y^x$ diverges as $|E_g|^{-1/2}$.

\section{Model Hamiltonians of essentially noncentric insulators}\la{sec:models}

Beside  corroborating propositions (P1-2,Q1-2), the models below are meant to illustrate the complementary roles of the intra-band Berry-Zak phase and the inter-band optical vorticity in determining the intercellular shifts, as well as to give intuition on the type of tight-binding hoppings that result in a shift obstruction.  One potentially surprising finding is that nontrivial optical vorticity (with a trivial Berry-Zak phase) leads to a large frequency-integrated shift conductivity, despite the shift connection vanishing at the $\bk$-position of the optical vortex.  Special attention is focused on identifying non-analyticities of shift-related quantities at various types of topological phase transitions.

\subsection{Model with second-class phase transition}\la{sec:modelsecondclass}

\subsubsection{Flatband limit with zero optical vorticity}

To realize a simple 2D model Hamiltonian with a reverting Thouless pump, I begin with the standard parametrization of a real-valued, unit-norm three-vector by two  spherical angles: $\bd{=}[\sin(\theta)\cos(\phi),\sin(\theta)\sin(\phi),\cos(\theta)]$, then replace $(\theta,\phi)$ by dimensionless wavenumbers $(q_x,q_y){:}{=}(k_xR_x,k_yR_y)$ and define the Hamiltonian $h(\bk){=}\bd(\bk){\cdot}\bsigma$. Take special note of the replacement of  $\theta{\in}[0,\pi]$ with  $q_x{\in} [-\pi,\pi]$. The motivation for this strange construction of the Hamiltonian is now evident: $\bd(\bk)$ covers the unit-norm sphere as $\bk$ is varied over $BZ/2$; this covering happens again (but with opposite orientation) over the other half of the BZ. Applying Berry's relation between the Berry curvature and the solid angle subtended by $\bd(\bk)$,\cite{berry_quantalphase} I establish that ${RTP}_v{=}1$. This can be alternatively established by computing the Zak phase as ${Z}_v{=}\pi[1{-}\cos(q_x)]$.\\

By construction,  the energy gap (separating flat conduction and valence bands) equals $2||\bd(\bk)||{=}2$, which defines the energy scale for my dimensionless Hamiltonian. One may verify the forementioned reflection symmetry of the Hamiltonian, as well as a time-reversal constraint $\hat{T}h(\bk)\hat{T}^{-1}{=}h(-\bk)$ with $\hat{T}{=}\sigma_3K$. The Fourier transform of $h(\bk)$ gives a real-space-dependent Hamiltonian with two intra-orbital hoppings over $(x,y){=}(R_x,0)$, and one inter-orbital hopping over $(R_x,R_y)$; the presumed insignificance of hoppings parallel to the polar axis ($y$) requires that the Wannier orbitals are highly anisotropic. This requirement is  in line with expectations that ideal shift-current materials necessarily have strongly delocalized and highly anisotropic covalent bonds.\cite{youngrappe_firstprinciples,LiangTan_upperlimit} It is hoped that the simplicity of my model (having only three independent hoppings) offers a generalizable insight to the type of covalent bonding that is conducive to shift currents.   \\

\begin{figure}[h]
\centering
\includegraphics[width=5 cm]{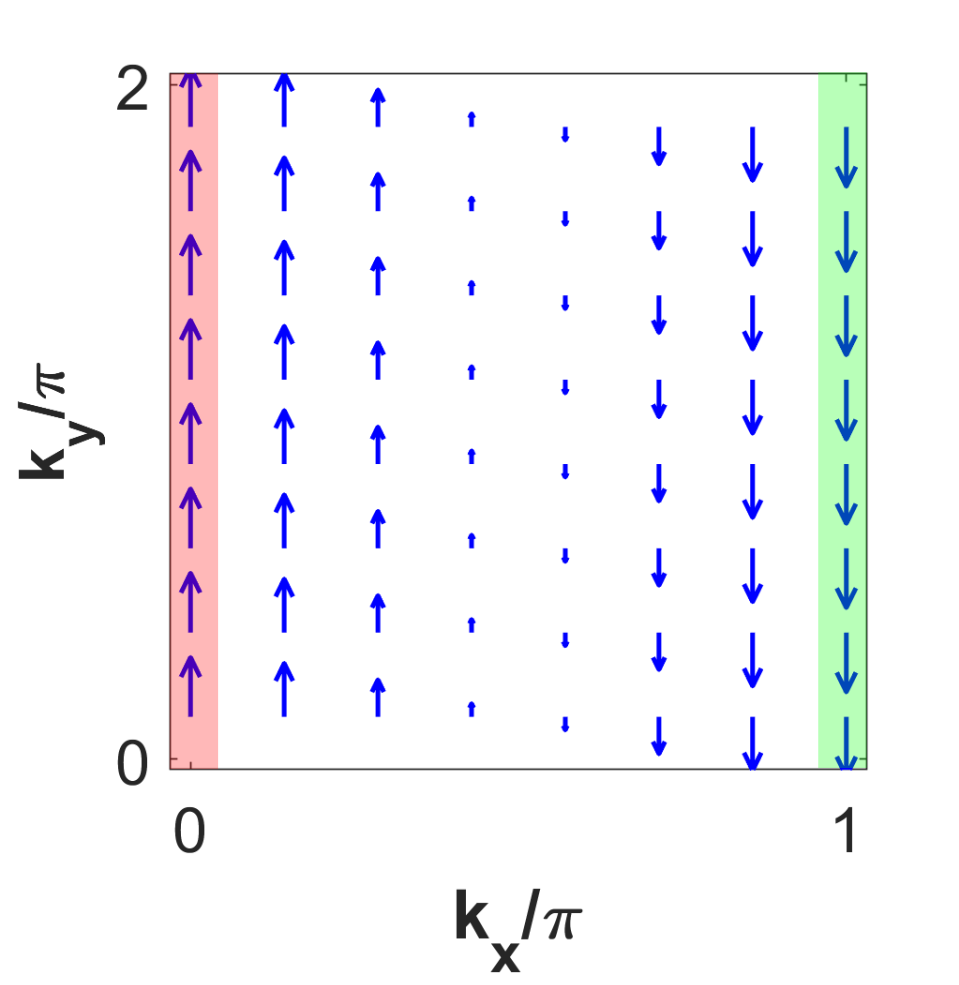}
\caption{Shift vector field for the flatband model, with lattice constants set to one. The vector field is plotted over $BZ/2$; the field over the other half of the Brillouin zone is fixed by time-reversal symmetry. Averaging the shift vector over either reflection-invariant $\bk$-line (colored red and green) gives the intercellular shift. }\label{fig:shiftvectorfield_flatband}
\end{figure}

While flat bands are often associated to atomic insulators with a trivial shift connection, the flat bands in my model arise purely from the inter-site hopping matrix elements, and the shift connection can be calculated as: \e{ {C}^{x}_{ycv\bk} \eq - \f{\eps_{\ab\delta}}{4}n_\alpha(\partial_{k_y}\partial_{k_x} n_\beta)(\partial_{k_x} n_\delta)=\bigg[\f{R_x^2}{4}\bigg]\bigg[R_y\cos(q_x)\bigg],\notag}
with all indices on the Levi-Cevita tensor contracted with indices on $\bn{:}{=}\bd/||\bd||$.  The $\bn$-vector expression for the shift connection manifests its sole dependence on the wave function, i.e., the position on the Bloch sphere. The quantity in the first [resp.\ second] square bracket is identifiable with $|{A}_{xcv\bk}|^2$ [resp.\ with $S^x_{ycv\bk}$]. Because of the $\bk$-independence of the optical affinity, there are no optical vortices: ${Vort_x}{=}0$. Integrating $S^{x}_{ycv\bk}{=}R_y\cos(q_x)$ over each reflection-invariant $\bk$-line [cf.\ \q{averageshift} and \fig{fig:shiftvectorfield_flatband}] gives the intercellular shifts as ${\scrs_0}{=}1{=}{-}{\scrs_{\pi/R_x}}$; the last equality, in combination with the previously-established ${RTP}_v{=}1$, establishes agreement with the shift obstruction relation [\q{shiftobstruction}]. While the BZ-integral of ${C}^{x}_{ycv}$ vanishes, ${C}^{x}_{ycv}$ has a peak with maximum value $R_x^2R_y{\scrs_0}/4$ at $k_x{=}0$, corroborating an estimate made in \s{sec:theory}. We see as a matter of principle that large values of the shift connection are attainable even if the band gap is infinitely larger than the band width: $E_g/E_w{=}\infty$. 

\subsubsection{Deviating from the flatband limit}

To demonstrate the robustness of the above topological invariants, I  introduce nearest and next-nearest inter-orbital hoppings in the $x$ direction, which corresponds to the Hamiltonian term: $\delta h(\bk){=}[\alpha\sin(q_x){+}\beta \sin(2q_x)]\sx$. Let us first consider a small deviation from the flatband limit, such that we remain in the same topological phase. The shift vector fields for $(\alpha,\beta){=}(1/5,-1/4)$ and $(1/4,1/8)$ are illustrated in \fig{fig:shiftvectorfield}(a) and (b) respectively. These picture panels may be compared to the flatband limit in \fig{fig:shiftvectorfield_flatband}. We see that the vector fields are continuously deformable but maintain a certain rigidity: the average of the shift vector over each reflection-invariant $\bk$-line is invariant. \\

\begin{figure}[h]
\centering
\includegraphics[width=8.6 cm]{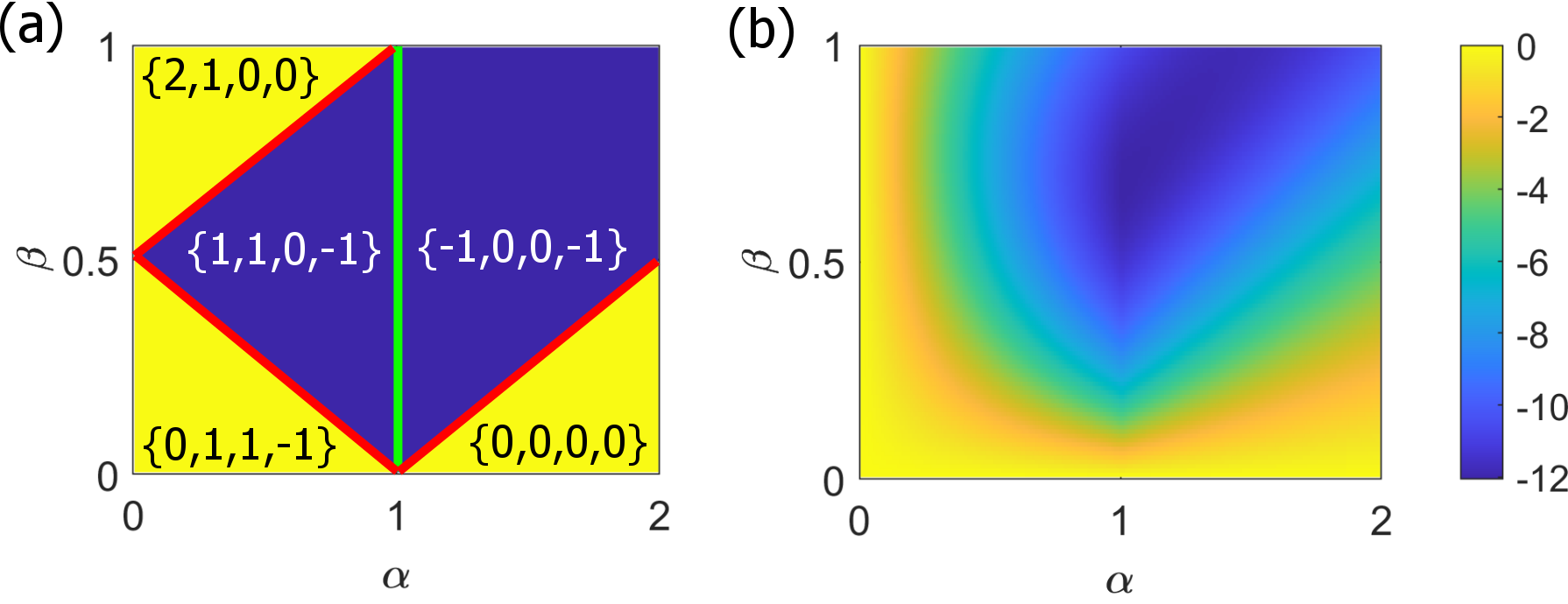}
\caption{(a) Phase diagram, with each phase is labelled by the invariants: $\{-Vort_x,RTP_v,\scrs_{0},\scrs_{\pi/R_x}\}$. The yellow region indicates the average intercellular shift $\langle \scrs \rangle{=}0$; the dark blue region indicates $\langle \scrs \rangle{=}{-1/2}$.  (b) Figure of merit ($F^x_{y,2D}$) corresponding to the phase diagram, with the numerical value for $F^x_{y,2D}$ indicated by a color bar.}\label{fig:alphabeta}
\end{figure}

With larger values of $(\alpha,\beta)$, one can induce a topological phase transition so that  ${C}^{x}_{ycv}$ has a nonzero BZ-average.  The resultant phase diagram is shown in \fig{fig:alphabeta}(a), with each phase labelled by four integer invariants: $\{-Vort_x,RTP_v,\scrs_{0},\scrs_{\pi/R_x}\}$; the  phase-transition lines are of two types that we subsequently deal with in turn. 

\subsubsection{Optical phase transition} \la{sec:opticalphasetransition}

 The lines $\alpha{+}2\beta{=}1$, $\alpha{-}2\beta{=}1$ and $\alpha{-}2\beta{=}{-}1$ are colored red in \fig{fig:alphabeta}(a), and mark  \emph{optical phase transitions} where the energy gap remains nonzero but the optical affinity vanishes at the reflection-invariant wavevector $(k_x,k_y){=}(0,\pi), (\pi,\pi)$ and $(\pi,0)$, respectively. Approaching a generic point on an optical transition line, a pair of reflection-related optical vortices (with opposite circulation) are either nucleated or annihilated, depending on the direction in which one approaches the transition point. \\

\begin{figure}[h]
\centering
\includegraphics[width=8.6 cm]{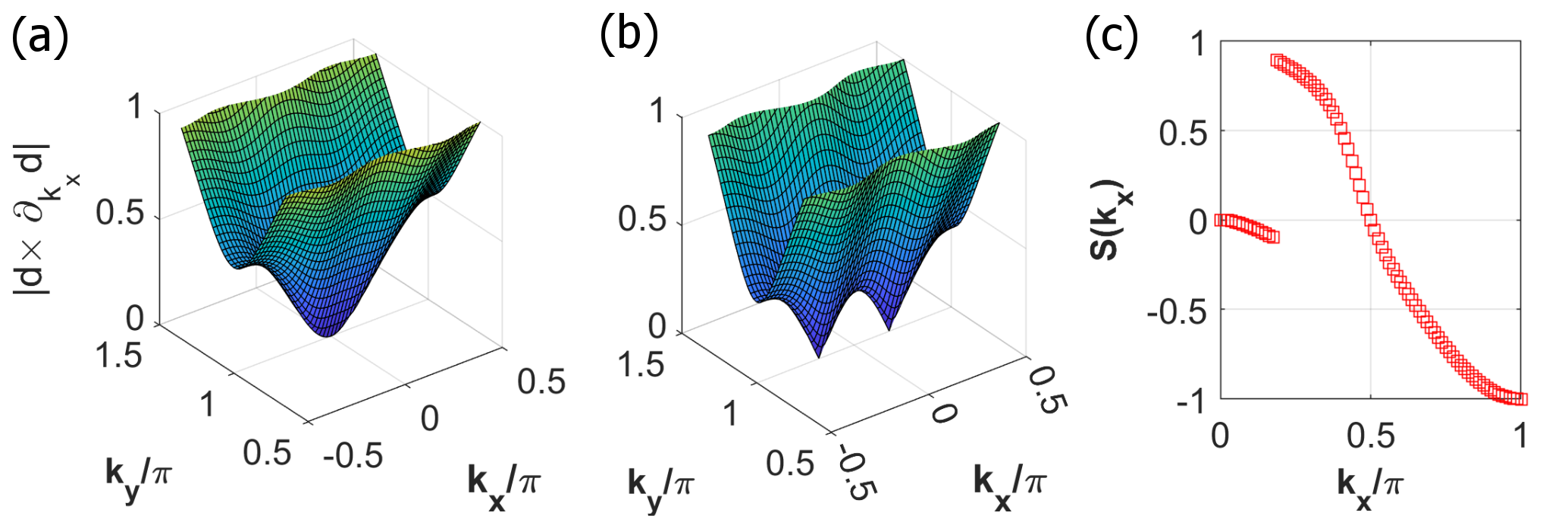}
\caption{(a-b) Zeroes of $||\bd\times \partial_{k_x}\bd||$ reveal optical zeroes, with $\bd$ the vector in the Hamiltonian $h(\bk){=}\bd(\bk){\cdot} \bsigma {+}h_{id}(\bk)I_{2\times 2}$. The model parameters are $\alpha{=}2\beta{=}1/2$ and $5/8$, respectively, for panels (a) and (b). Panel (c) plots the line-averaged shift $\scrs(k_x)$ [\q{averageshift}] for $\alpha{=}2\beta{=}5/8$. All lattice periods have been set to one.}\label{fig:vorticity}
\end{figure}

To visualize this process, I employ the Hamiltonian-vector interpretation of optical vortices [introduced near \q{Acvidentity}] to track the $\bk$-locations of optical zeroes and vortices -- by plotting $||\bd{\times}\partial_{k_x}\bd||$ over the BZ. For instance, increasing $\alpha{=}2\beta{=}0$ from zero, the minimal optical affinity vanishes at the optical phase transition $\alpha{=}2\beta{=}1/2$, as illustrated in \fig{fig:vorticity}(a); this  optical zero has vanishing circulation, but can be interpreted as the merging of a pair of optical vortices with cancelling circulations. Indeeed, by further increasing  $\alpha{=}2\beta$ to $5/8$, the pair of vortices split away in \fig{fig:vorticity}(b). \\

Each of the vortices manifests as a circulation in the shift vector field [illustrated in \fig{fig:shiftvectorfield}(c)] as well as a unit discontinuity in the $k_x$-dependent line-averaged shift [illustrated in \fig{fig:vorticity}(c)]. On the other hand, the intra-band-Berry-phase invariant $RTP_v$ is unchanged across an optical phase transition, because the energy gap does not vanish.\footnote{For an insulator with trivial Chern invariants, the existence of wave functions which are analytic (with respect to $\bk$) and periodic over the Brillouin torus is guaranteed, which implies the intra-band Berry connection is also analytic and periodic. See references for the Grauert-Oka theorem in footnote 12 of \ocite{nogo_AAJH}.} The invariance of $RTP_v$ and the unit change in optical vorticity jointly imply that the relative intercellular shift must change by one unit, according to the shift obstruction relation [\q{shiftobstruction}].  

\subsubsection{Energetic phase transition}

The $(\alpha{=}1)$ line is colored green in \fig{fig:alphabeta}(a), and marks an \emph{energetic phase transition} where the energy gap closes at only  two non-symmetric wavevectors: $\bq{=}({\pm}\pi/2,\pi)$. For any point in the left half of the phase diagram ($\alpha{<}1$), $RTP_v{=}1$ is deducible by energy-gap-preserving continuity to the flatband limit: $(\alpha,\beta){=}(0,0)$. Across the $\alpha{=}1$ line, the Zak phase ${Z}_v(\pi/2)$ changes discontinuously by $\pi$, resulting in $RTP_v{=}0$ for $\alpha {>}1$.\\

To understand the $\pi$ discontinuity, consider that the $k_y$-dependent Hamiltonian at fixed $q_x{=}\pi/2$ (and for any value of $\beta$) has a Hamiltonian vector with components:  $d_1{=}\cos q_y{+}\alpha,\;d_2{=}\sin q_y, \;d_3{=}0$. Viewing $(d_1,d_2)$ as a two-vector on a plane, the two-vector makes one full revolution around the origin as $q_y$ is advanced by $2\pi$, if $|\alpha|{<}1$. Otherwise, no net revolution is made. This discontinuity in revolution number manifests as the Zak phase equalling $\pi$ for  $|\alpha|{<}1$, and equalling zero otherwise. This discontinuous change in the Zak phase (at $q_x{=}\pi/2$) converts a reverting Thouless pump [illustrated in \fig{fig:zakphase}(a)] to a trivial pump [\fig{fig:zakphase}(b)].   \\

 A unit change in ${RTP}_v$ (that arises from a band touching at a non-symmetric wavevector) implies that the optical vorticity must change by two units, so as to satisfy the shift obstruction relation.\footnote{The relative intercellular shift is invariant across $\alpha{=}1$;  $\Delta \scrs$ can only change if either the energy gap or optical affinity vanishes at a mirror-invariant $\bk$, as elaborated in \app{app:invariants}.} How the optical vorticity changes by two (across $\alpha{=}1$) is a process of \emph{vorticity inversion}: an optical vortex is `swallowed' (at the band touching point) then `spat out' with opposite circulation. \\

To understand this inversion, we return to the Hamiltonian-vector interpretation: recall that an optical vortex is an optical zero with nontrivial circulation, and an optical zero is a wavevector $\bk_0(\alpha)$ where $\bd\times \partial_{k_x}\bd{=}0$. $\alpha{=}1$ marks a transition where $\bd$ and $\partial_{k_x}\bd$ change from being parallel to antiparallel; this is possible because $\bd(\bk_0(\alpha))$, being proportional to the energy gap, vanishes at the transition point. The inversion in the orientation of $\bd$ implies that, as $\bk$ is advanced in a small circle around $\bk_0$, the sense of relative rotation (between $\bd$ and $\partial_{k_x}\bd$) is also inverted -- this is why the optical vortex flips its circulation.  

\subsubsection{Figure of merit}\la{sec:figureofmerit}

 Over the same  range for the Hamiltonian parameters $(\alpha,\beta)$, \fig{fig:alphabeta}(b) shows a numerically generated  plot of the dimensionless figure of merit $F^x_{y,2D}{:}{=}(2\pi/R_x)\int_{BZ}d^2k {C}^{x}_{ycv}$, which is the 2D analog of $F^x_{y}$ in \q{bandshift}. The numerical value for $F^x_{y,2D}$ indicated by a color bar on the right of the figure panel.  Comparison of panels (a) and (b) in \fig{fig:alphabeta} reveals: \\

\noi{i} A positive correlation of $F^x_{y,2D}$ with the average intercellular shift $\langle \scrs\rangle$; the latter quantity vanishes in the yellow region of  \fig{fig:alphabeta}(a), and equals $-1/2$ in the dark blue region. \\

\noi{ii} Phases with different  $\langle \scrs\rangle$ are separated by optical phase transitions [indicated by red lines in \fig{fig:alphabeta}(a)]. Suppose one defines a trajectory on the phase diagram starting from $\langle \scrs\rangle{=}0$ (yellow region) and ending at $\langle \scrs\rangle{=}{-}{1/2}$ (dark blue region), there is a continuous crossover in the value of $F^x_{y,2D}$ from $0$ to about ${-}10$, if the trajectory does not start or end too close to an optical transition line. If the same trajectory does not intersect an energetic transition line (colored green), then the crossover (of $F^x_{y,2D}$) is not just continuous but smooth. One can verify the smoothness by asymptotic analysis: fixing $\alpha{=}2\beta$ and parametrizing the approach to the optical transition line ($\alpha{+}2\beta{=}1$) by a new variable $\delta{=}\alpha{-}1/2$; one finds that the shift vector diverges as $1/\delta$, but this divergence is cancelled by the vanishing of the optical affinity: $|{A}_{xvc}|^2{\propto} \delta^2$.  \\

\begin{figure}[h]
\centering
\includegraphics[width=8.6 cm]{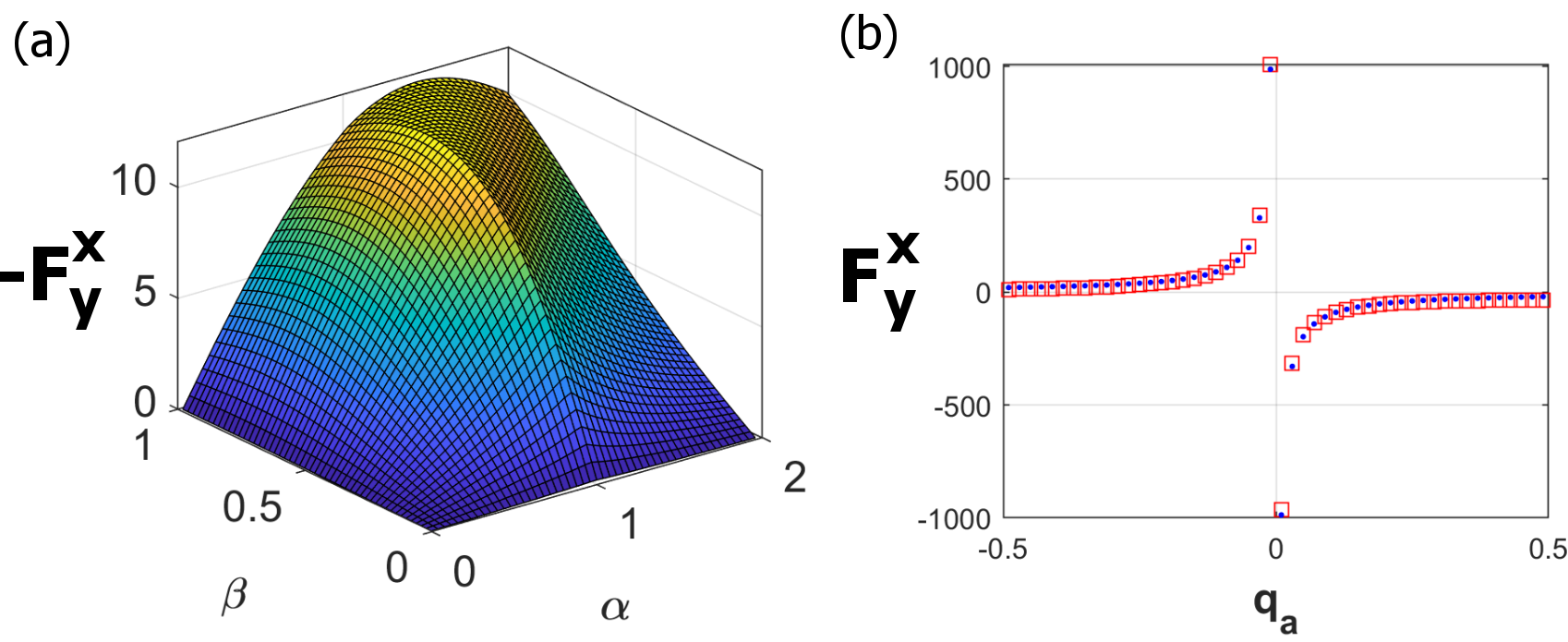}
\caption{(a) Kink-type non-analyticity of the figure of merit $F^x_{y,2D}$, for a second-class phase transition. (b) Divergent non-analyticity for a first-class phase transition modelled in \s{sec:modelfirstclass}: red squares represent a numerical integration, and blue dots represent an analytically-derived formula: ${-}\pi^2/q_a$.}\label{fig:nonanalytic}
\end{figure}

\noi{iii} In contrast, there is a non-analyticity of $F^x_{y,2D}$ across the energetic phase transition (green line, $\alpha{=}1$), because both the shift vector and the optical affinity  diverge. For a quantitative analysis, let me introduce a new variable $Q$ by $Q{+}1{=}\alpha{=}2\beta$. One finds  that $|Q|$ is simply the minimal energy gap over the BZ, $S^x_{ycv}$  diverges as $|Q|^{-1}$, and $|{A}_{xvc}|^2$ diverges as $Q^{-2}$; the latter observation implies that the phase transition is second-class, according to the classification made in \s{sec:implications}. By dimensional analysis, one deduces that $F^x_{y,2D}$ equals the sum of an analytic function of $Q$ plus a non-analytic power series: $a_1/|Q|{+}a_2\sgn[Q]{+}a_3|Q|{+}\ldots$. For this model, one can prove $a_1{=}a_2{=}0$,\footnote{This is partially understandable from the shift connection being odd under $(\delta q_x{+}Q/2){\ri}(-\delta q_x{+}Q/2)$ for sufficiently small $|Q|$ and $|\delta q_x|$; here, $\delta q_x{=}q_x{-}\pi/2$ is the wave number measured from the point of closest, inter-band contact.}  leading to a kink-type non-analyticity which is faintly visible in \fig{fig:alphabeta}(a) as a darkening localized to the  $(\alpha{=}1)$-line, but is more evident in \fig{fig:nonanalytic}(a) where $F_y^x(\alpha,\beta)$ is shown as a three-dimensional surface plot.\\

\noi{iv} Recall an earlier observation that  $F^x_{y,2D}{\approx}{-}10$ in the trapezium-shaped phase on the right corner of  \fig{fig:alphabeta}(a). This phase represents an insulator with trivial Chern number and trivial reverting pump ($RTP_v{=}0$); overall, this insulator would be considered trivial by the standard classification of topological insulators based on the intra-band Berry connection. The largeness of $|F^x_{y,2D}|$ is thus solely attributed to the nontrivial inter-band optical vorticity ($Vort_x{=}1$). This attribution may  surprise some readers, because the existence of an optical vortex implies that the optical affinity (hence also the shift connection) vanishes at the $\bk$-position of the vortex. However such vanishing occurs only in a measure-zero $\bk$-region with codimension two, i.e., only at isolated points in a 2D BZ. There is a competing and manifestly dominant factor: the vortex induces large variations of the shift vector over half the BZ period (according to the shift-obstruction relation), resulting in the average intercellular shift being ${-}{1/2}$ and $|F^x_{y,2D}|{\gg}1$.\\

 One may view $h(\bk){+}\delta h(\bk)$ as a $k_z$-independent Hamiltonian for a 3D insulator that is constructed by stacking many layers of the 2D insulator with weak inter-layer coupling. Then the 2D figure of merit (of the 2D insulator) is simply proportional to the 3D figure of merit (of the 3D layered insulator):  $F^x_{y,2D}{=}(R_x/R_z)F^x_{y}$, with a proportionality factor that is a ratio of lattice periods and is typically ${\approx}1$. Then the observations made in (i-iv) above have 3D analogs which support proposition (Q1). In particular, $F_y^x{\approx}{-10}$ with $\langle \scrs\rangle{=}{-}{1/2}$ is just slightly larger than an order-of-magnitude estimate ($F_y^x {\sim} 2\pi \langle\scrs\rangle R_x/R_z$) made in \s{sec:theory}. However, because the leading non-analyticity of $F^x_{y}$ is of the kink-type: ${\sim}|E_g|$, proposition (Q2) does not apply to the second-class phase transition of this model.  

\subsection{Model with first-class phase transition} \la{sec:modelfirstclass}

To have $F^x_{y}$ diverge as $|E_g|^{-1/2}$ [proposition (Q2)], I offer a different model Hamiltonian: $h(\bk){=}{-}(\dg{z}\bsigma z){\cdot} \bsigma$ with
\e{	 z(\bk)=\vectwo{z_1}{z_2}=\vectwo{\sin q_x}{\sin q_y+i(q_a+\sum_{j=x,y}\cos q_j-2)}. \la{firstclassmodel}}
$(q_x,q_y){=}(k_xR_x,k_yR_y)$ are dimensionless wavenumbers, and $q_a$ is a real-valued tuning parameter that induces the band gap (${=}2\dg{z}z$) to close when $q_a{=}0,2$ and $4$; there are no optical transitions induced by $q_a$. I will focus on the $q_a{=}0$ transition where the band gap closes at $\bk{=}(0,0)$;
an effective, low-energy Hamiltonian  describing the transition is obtained by truncating the Taylor expansion of $z(\bk)$ with respect to $\bk$:  
\e{ h_{t}(\bk)=-(\dg{z}_t\bsigma z_t)\cdot \bsigma, \as z_t(\bk)=\vectwo{q_x}{q_y+iq_a}.}
Reflection and time-reversal symmetries manifest as $\sigma_3h(\bk)\sigma_3{=}h(-k_x,k_y)$ and $h(\bk){=}\overline{h(-\bk)}$, respectively.\\

  The form of the Hamiltonian is inspired by previous models of reverting Thouless pumps that are protected by a different crystallographic symmetry: rotation.\cite{AA_teleportation,nelsonAA_multicellularity,nelsonAA_delicatetopology} By design, ${RTP}_v$ changes by unity across $q_a{=}0$, which may be understood from a $2\pi$-discontinuity of the Zak phase ${Z}_v(k_x{=}0)$ at a reflection-invariant $\bk$-line. (This contrasts with the $\pi$-discontinuity of the Zak phase at a non-symmetric $\bk$-line studied in the previous second-class phase transition.) A rough understanding of the $2\pi$-discontinuity follows from inspecting the normalized, valence-band eigenvector solution to $h_t(\bk)$: $\ket{{u}_v(\bk)}{=}(-q_y{+}iq_a,q_x)/\sqrt{\dg{z}_tz_t}$, and realizing that the phase of  $\ket{{u}_v(0,k_y)}$ changes by $2\pi$ as $(q_y,q_a)$ is varied over a  circle with radius $(q_y^2{+}q_a^2)^{1/2}$.\footnote{For a more direct proof, consider that the valence-band eigenvector solution to $h(\bk)$ is $\ket{u_v(\bk)}{=}(-\overline{z_2},z_1)/\dg{z}z$, which is practically unchanged as one tunes $q_a$ across zero, except for $\sqrt{q_x^2{+}q_y^2}$  small enough  to be comparable to $|q_a|$. This implies that the $2\pi$-discontinuity of the Zak phase ${Z}_{v,0}$ [for $h(\bk)$] can be derived from a $2\pi$-discontinuity of the continuum analog of the Zak phase: ${Z}_v^{ctm}{=}\int_{-\infty}^{\infty}{A}_{yvv(0,k_y)}dk_y$, with the Berry connection ${A}_{yvv}$ a functional of the eigenvector  solution of $h_t(\bk)$. This solution being simply $({-}q_y{+}iq_a,q_x)/||\bq||^2$, one deduces ${Z}_v^{ctm}{=}\pi \sgn[q_a]$, as desired.} For $|q_a|{\gg}1$, one deduces ${RTP}_v{=}0$ from the simple form of the Hamiltonian $h(\bk){\approx}q_a^2\sz$; thus it must be that ${RTP}_v{=}{-}1$ for $q_a{\in}(0,2)$.  \\

What of the optical vortices? For large $|q_a|$, there are four vortices positioned at wavevectors $(q_x,q_y){\approx} ({\pm}\pi/2,0)$ and  $(\pm\pi/2,\pi)$, with small corrections (of order $1/q_a$) to the $q_x$-component of these positions. The two vortices in $BZ/2$ have opposite circulation, hence the vorticity invariant ${Vort_x}$ vanishes. As $q_a$ approaches $0$ from the negative side, two of the four vortices merge at the band-touching point and then mutually annihilate, leaving behind a net vorticity ${Vort_x}{=}{-}1$ for $q_a{\in}(0,2)$. \\

Having determined ${RTP}_v$ and ${Vort_x}$, the shift obstruction relation tells us that the relative intercellular shift $\Delta \scrs$ vanishes for $q_a{<}0$ and equals ${+}1$ for $q_a{\in}(0,2)$. Because of the integer-quantization of $\scrs_{0}$ and $\scrs(\pi)$, the parities of $\Delta \scrs$ and $\scrs_{0}{+}\scrs(\pi){=}2\langle \scrs\rangle$ must equal. A calculation gives explicitly that $\langle \scrs\rangle{=}0$ for $q_a{<}0$ and  $\langle \scrs\rangle{=}{-}1/2$ for $q_a{\in}(0,2)$. At the mid-point ($q_a{=}1$) between two energetic phase transitions, I numerically evaluate  $F^x_{y,2D}{=}(R_x/2\pi)\int C^{x}_{ycv}d^2k{\approx}{-39}$, which is a factor of four larger than the analogous value for the previous model [\s{sec:figureofmerit}].  \\

The energetic phase transition is accompanied by the optical affinity diverging as $q_a^{-2}$, and the shift vector diverging as $q_a^{-1}$. Identifying $2\dg{z}{z}|_{\bk{=}\bze}{=}2q_a^2$ as the minimal band gap $E_g$, we deduce that the phase transition is first-class.
From asymptotic analysis,  $F^x_{y,2D}{\approx} c_1/q_a {\approx} c_1(2/E_g)^{1/2}$ for sufficiently small $|q_a|$, with $c_1$ a dimensionless constant. One can analytically evaluate $c_1{=}{-}\pi^2/\sqrt{2}$, which is confirmed also by a numerical integration in \fig{fig:nonanalytic}(b).\\

 Finally, if we view $h(\bk)$ as a $k_z$-independent Hamiltonian for a 3D insulator, then the divergence of the 2D figure of merit also applies to the 3D figure of merit: $F^x_{y}{=}{(R_z/R_x)}F^x_{y,2D}$, giving us proposition (Q2); one deduces also that $F^x_{y}{\approx}{-39(R_z/R_x)}$ for $\langle \scrs\rangle{\approx}{-}1/2$, in support of proposition (Q1).

\section{Discussion and outlook}\la{sec:discussion}

The well-known topological insulators (e.g., the Chern\cite{Haldane1988} or $\Z_2$ topological insulators\cite{kane2005A,fukanemele_3DTI,moore_3DTI,Rahul_3DTI}) are compatible with having a center of inversion, and hence compatible with a zero bulk photovoltaic current. There exists a less-known class of topological insulators which are \emph{essentially noncentric}, meaning that the topologically nontrivial phase of matter exists only in crystal classes without a center of inversion. This work was motivated by the question of whether essentially noncentric topological insulators can have large excitation shift currents with large band gaps. This work establishes an affirmative answer for a subset of essentially noncentric insulators that are polar/pyroelectric. \\

There has been a fruitful tradition of identifying what properties a topological insulator absolutely cannot have, e.g., zero quantum entanglement,\cite{turner_entanglementinversion,hughes_inversionsymmetricTI,aa_traceindex,ZhoushenHofstadter} analytic Bloch functions,\footnote{For related references, see \ocite{Panati_trivialityblochbundle} and footnote 12 in \ocite{nogo_AAJH}} symmetric Wannier functions which are localized to various degrees,\cite{Brouder_explocWFs,TQC,nogo_AAJH,crystalsplit_AAJHWCLL,nelsonAA_multicellularity,read_compactwannier} trivial Berry-Zak phase.\cite{fukane_trspolarization,yu_equivalentZ2,AA_wilsonloopinversion,crystalsplit_AAJHWCLL,bouhon_wilsonloopapproach} This work demonstrates that some essentially noncentric insulators are characterized by a \emph{shift obstruction}: the inability to continuously tune the photonic shift vector to zero throughout the Brillouin zone (BZ). This obstruction depends on the difference between an intra-band-Berry-phase invariant (the reverting Thouless pump) and an inter-band-Berry-phase invariant (the optical vorticity), as shown in \q{shiftobstruction} for two-band Hamiltonians, and in \q{Nbandshiftobstruction} for $(N{>}2)$-band Hamiltonians.\\

The shift obstruction exemplifies a new class of topological invariants that depend on both the intra- and inter-band Berry connections; by `inter-band', I mean the connection between valence and conduction bands.\footnote{My emphasis on topological aspects of the inter-band Berry connection is philosophically akin to a recent Riemannian-geometrical interpretation of the dipole matrix element\cite{ahn_riemanniangeometry}}  One implication is that the topological theory of nonlinear optical responses does not reduce or simplify to the standard theory of  topological insulators; this standard theory is based on the characterization of the intra-band Berry connection but not the inter-band Berry connection. Topological insulators which are  trivial in the standard classification can have nontrivial invariants in the `\emph{opto}pological' classification presented here. This  classification is demonstrated in \app{sec:beyond2band} to be an optopological generalization of  `symmetry-protected delicate topology',\cite{nelsonAA_multicellularity,nelsonAA_delicatetopology} in the sense that the meaning of a topological invariant defined for a two-band Hamiltonian can be extended to an $(N{>}2)$-band Hamiltonian, subject to conditions on the symmetry representations of all $N$ bands.

\subsection{Experimental implication: transient vs steady photovoltaic current} \la{sec:transient}

A nontrivial shift obstruction generically implies a large frequency-integrated excitation shift conductivity ($\sigma^{\text{exc},j}_{i}$); this is supported by a plausibility argument [\s{sec:theory}] and model calculations [\s{sec:models}]. Largeness has been quantified by a figure of merit $F^{j}_i$, that we define as the BZ-integral of the photonic shift connection in \q{bandshift}. \q{zeroT} shows  $2 F^{j}_i \f{e^3}{h^2}$ to be the frequency-integrated shift conductivity due to inter-band photoexcitation of a zero-temperature insulator, assuming that excitons are weakly bound.\cite{morimoto_excitonic}  \\

As explained in a companion paper,\cite{AAzhu_anomalousshift}  the excitation shift current associated to $\sigma^{\text{exc}j}_{i}$ is the \emph{transient photovoltaic current} that follows the onset of radiation. Specifically, setting time $t{=}0$ at the onset, we consider the photovoltaic current at $t{<} \tau_{e-p}$, with $\tau_{e-p} {\sim} 100  fs$ a typical time scale for electron-phonon collisions. In this early time regime, the photoexcited electron-hole system has not relaxed (within a band) or recombined (across the band gap), thus the transient current is essentially the excitation shift current.\cite{AAzhu_anomalousshift} The transient photocurrent (${\propto}\sigma^{\text{exc}j}_{i}$) may either be measured directly with an ultrafast oscilloscope (with sub-picosecond resolution) or indirectly by measuring the emitted radiation induced by pulsed photoexcitations.\cite{laman_ultrafastshift,braun_ultrafastphotocurrents,sotome_spectraldynamics} \\

The  shift obstruction relation [\q{shiftobstruction}] implies that optical vorticity can induce a large transient shift current. There are two competing effects of vorticity: while it is well-known that the photonic shift connection vanishes \emph{locally} at the $\bk$-position of the vortex center, vorticity also induces large variations of the photonic shift vector over the scale of the BZ period (according to \q{shiftobstruction}), suggesting plausibly that the \emph{momentum-integrated} shift connection is large. A model calculation in \s{sec:modelsecondclass} identifies this BZ-wide shift-vector variation as dominating over the local vanishing of the shift connection. Thus if one is interested in inducing a large transient photovoltaic current by a \emph{broadband} light source, even materials with a trivial intra-band Berry phase (i.e., negligible polarization) may be looked upon as favourable candidates -- if they have nontrivial optical vorticity.  \\

While the above argument for vorticity-induced shift variations has been verified for essentially noncentric insulators, actually the argument more generally applies to any insulator with optical vorticity; indeed, any insulator can host stable optical vortices, because the robustness of optical vortices depends only on the discrete translational symmetry, and not on any crystallographic point-group symmetry.\footnote{In this respect, optical vortices are the optopological analogs of Weyl points in topological (semi)metals.\cite{Nielsen_ABJanomaly_Weyl,wan_weylsemimetal,halasz_weylsemimetal}} \\

Beyond early-time transient behavior, the steady photovoltaic current comprises not just the excitation shift current, but also includes: (i) additional components of the shift current due to inter-band recombination  and intra-band relaxation,\cite{belinicher_kinetictheory} which have their own wave-function-geometric interpretation;\cite{AAzhu_anomalousshift} (ii) a non-shift (`ballistic') contribution,\cite{belinicher_ballistic} which originates from an  asymmetry of the quasiparticle distribution ($f(\bk){\neq}f(-\bk)$) induced by intra-band scattering,\footnote{A large shift current does not necessarily imply a large ballistic current. A typical peak value for the frequency-dependent ballistic conductivity is $30 \mu A/V^2$ in magnitude;\cite{sturman_ballisticandshift,alperovich_gaas,zhenbang_phononballistic} which is small compared to the large shift conductivity we predict. } and (iii) a photon-dragged current that is entrained to the photon momentum.\cite{danishevskii_dragging,grinberg_lightpressure} The full impact of optical vorticity on the steady photovoltaic current has not been elucidated, but it is now apparent that the optical vorticity results in the steady shift current being highly sensitive to the light polarization.\cite{AAzhu_anomalousshift}\\

The topological perspective of the shift vector potentially has utility beyond the bulk photovoltaic effect. For instance, the shift vector also plays an important role in second harmonic generation,\cite{sipe_secondorderoptical,morimoto_nonlinearoptic,Panday_strong2HG} and in ultrafast optical rectification for frequencies above the band gap.\cite{nastos_opticalrectification} The latter effect emits THz radiation that is desirable for spectroscoscopy.

\subsection{Outlook for material searches}

Ab-initio-based, high-throughput searches for topological materials have largely focused on band inversion as a diagnosis criterion for being topologically nontrivial.\cite{TQC,tiantian_catalogue,Maia_completecatalog,tang_symmetryindicators} To clarify the meaning of `band inversion', there exists for trivial insulators\footnote{In this context, a trivial insulator has a valence subspace that is a band representation.\cite{crystalsplit_AAJHWCLL}} a natural ordering (on the energy axis) of the representations of certain crystallographic point-group symmetries, and for nontrivial insulators this ordering is inverted. For instance, if the rotational (resp. parity) representations are inverted, one is guaranteed to have a topological Chern insulator (resp. $\Z_2$ topological insulator);\cite{Chen_bulktopologicalinvariants,Fu_inversionsymmetry} neither of these insulators is {essentially noncentric}, and therefore each is compatible with a zero bulk photovoltaic current. In contrast, our newly-introduced class of essentially noncentric topological insulators  are not band-inverted, which may be verified from  the model Hamiltonians in \s{sec:models}, as well as model extensions described in \app{sec:extensions}.\footnote{Being un-inverted and still topologically nontrivial occurs for some `fragile' topological insulators\cite{crystalsplit_AAJHWCLL,fu_tci,spinorbitfree_AAchen} and all known `delicate' topological insulators.\cite{nelsonAA_multicellularity,nelsonAA_delicatetopology,lapierre_nbandhopf}} \\

If not `band inversion', what serves as a diagnosis criterion for large  shift currents? One answer that was  proposed  in \ocite{fregoso_opticalzero} is to compute the inter-band polarization difference, assuming that optical vortices are absent. Such a computation requires to average the intra-band Berry phase over a reduced Brillouin zone,\cite{kingsmith_polarization} while fixing the phase of the wave function over the entire BZ in the `optical gauge',\cite{fregoso_opticalzero} which is a computationally expensive procedure. Moreover, if vortices were present, the inter-band polarization difference has questionable relevance to the shift current. It is therefore advantageous to directly relate the shift vector, the intra-band Berry-Zak phase and the optical vorticity on equal footing, without the ad hoc assumption that the vorticity vanishes. This relation is precisely given by the shift obstruction relation [\q{shiftobstruction}]. One lesson learned from this relation is that the inter-band polarization difference  is not a general criterion for large excitation shifts; largeness is generally attributed to an interplay of the intra- and inter-band Berry phases, with both quantities either competing or synergizing.

\subsubsection{Materials with nontrivial optical vorticity}

Given the prominent role played by optical vorticity in the transient and steady shift current [\s{sec:transient}], one would like to identify  nontrivial vorticity in a candidate noncentric material. This identification can be automated for a high-throughput ab-initio search.  Here is one possible algorithm: \\

\noi{a} Identify pairs of `optically-active' bands within an energy interval determined by the desired application, e.g., for solar-cell applications, the energy interval is determined by the solar spectrum. For each pair, ensure that one band lies in the valence subspace, and the other in the conduction subspace.\\

\noi{b} For each pair of optically-active bands labelled by $c$ and $v$, compute the affinity $|A_{jcv}|^2$ on a $\bk$-mesh over the Brillouin zone, for $j=x,y,z$. The affinity is calculable from existing ab-initio techniques\cite{youngrappe_firstprinciples,chong_firstprinciplesnonlinear,julen_wannierinterpolation,chongwang_nonorthogonal} with at least one of these techniques being fully automated for high-throughput calculations.\cite{chongwang_nonorthogonal}\\

\noi{c} For any $\bk$ on this mesh, if the affinity lies below a pre-decided threshold, perform a gradient-descent algorithm to determine if the affinity is reducible to zero (within some reasonable tolerance). \\

\noi{d} As an optional step to filter out false candidates, compute the photonic shift vector ($\bS^{\vec{\bj}}_{cv}$) on a $\bk$-mesh.  This vector diverges exactly at the vortex center, and will appear anomalously large for a $\bk$-point that is sufficiently close to the vortex center. The shift vector is also calculable from existing ab-initio techniques.\cite{youngrappe_firstprinciples,chong_firstprinciplesnonlinear,julen_wannierinterpolation,chongwang_nonorthogonal} \\

\begin{figure}[h]
\centering
\includegraphics[width=6 cm]{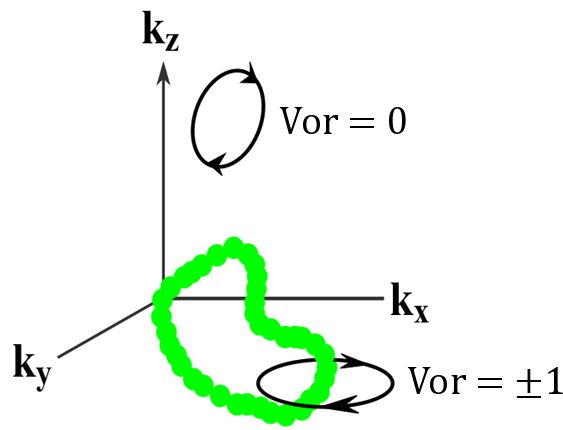}
\caption{In a three-dimensional Brillouin zone, optical vortices generically form lines. One representative line is colored green.}
\label{fig:vortexalgorithm}
\end{figure}

\noi{e} For the final test, compute 
\e{Vort^{\text{loop}}_j{=}{-}\oint \bS^{\vec{\bj}}_{cv\bk}{\cdot}\f{d\bk}{2\pi} \la{vortloop}}
over a small $\bk$-loop encircling the hypothesized $\bk$-location of the optical vortex,\footnote{\q{vortloop} reduces to the winding number of the inter-band Berry phase: $\oint \nabla_{\bk}\arg A_{jcv}{\cdot}d\bk/2\pi$ for an infinitesimal $\bk$-loop. For numerical simulations on practical $\bk$-meshes, one can disentangle inter- from intra-band contributions to \q{vortloop} by scaling the size of the $\bk$-loop; the intra-band contribution is proportional to the area enclosed by the $\bk$-loop (by Stokes theorem), while the inter-band contribution is insensitive to the size of the $\bk$-loop. } as illustrated in \fig{fig:vortexalgorithm}. The candidate fails the test if $Vort^{\text{loop}}_j{=}0$; for the generic optical vortex, $Vort^{\text{loop}}_j{=}{\pm}1$;  non-generic vortex with $Vort^{\text{loop}}_j{=}{\pm}n$ ($n{=}2,3,\ldots$) can also exist on  $\bk$-lines of high symmetry.   \\

Beyond high-throughput search algorithms, a Chern-vorticity theorem developed in \ocite{AAzhu_anomalousshift} predicts the existence of optical vorticity in topological semimetals, Chern insulators and insulators proximate to a trivial-$\Z_2$ topological phase transition; the latter is exemplified by the polar semiconductor BiTeI.\cite{AAzhu_anomalousshift}

\subsubsection{Essentially noncentric materials}

The present theory predicts large shift currents for essentially noncentric insulators  with the polar point groups $C_s$ and $C_n$ ($n{=}2,3,4,6$). Insulators within this subset of space groups should be filtered according to the symmetry representations of bands near the Fermi level [cf.\ \s{sec:beyond2band}, as well as discussions of the `mutually-disjoint' condition in \ocite{nelsonAA_delicatetopology}]. For candidate materials that survive filtration, I propose to compute the intercellular shift vector (or the generalized intercellular shift in \s{sec:beyondmonatomic}), which is an average of the photonic shift vector $S^j_{icv}$ over mirror- and/or rotation-invariant cross-sections of the BZ.\footnote{In contrast, the usual practice in the ab-initio community is to compute the $\bk$-dependent shift connection over the BZ and integrate the connection to obtain either $\sigma_{i\omega}^{\text{exc},j}$ or $\int\sigma_s^{abb}d\omega$. My proposal to compute the intercellular shift requires minimal modification of existing ab-initio packages, and merely redirects the spotlight to a different shift-related quantity defined over fewer $\bk$-points. }  For topologically nontrivial insulators with $C_s$ symmetry, the transverse intercellular shift was demonstrated here to be large, with the shift current parallel to a polar axis and the light polarization orthogonal to any polar axis. A further  calculation of the intraband Berry-Zak phase\cite{yu_equivalentZ2,z2pack} will reveal whether a shift obstruction (if present) derives from   a linear combination of the reverting Thouless pump and  optical vorticity, as per the shift obstruction relation [\q{shiftobstruction}]. \\

Future investigations will likely expand the list of noncentrosymmetric space groups that allow for essentially noncentric insulators with large shift currents. The existence of essentially noncentric topological insulators is known for other polar point groups (e.g., $C_{4v}$, $C_{6v}$\cite{spinorbitfree_AAchen}) as well as non-polar point groups;\footnote{A case in point is the Hopf insulator.\cite{Hopfinsulator_Moore,AA_teleportation} Most studied models of the Hopf insulator have a rotational axis, but this axis is removable because the Hopf invariant is well-defined without any point-group symmetry.} however, the shift current response for these insulators has never been been investigated. It is hoped that this work sparks the interest to do so.    \\

\noindent \textit{Post-submission addendum:} a subsequent work by Jankowski and Slager has demonstrated that the excitation shift conductivity (of the circular photogalvanic current), when integrated over frequency and suitably averaged over possible orientations of the current and electric field, is topologically quantized in certain models of antiferromagnetic insulators with neither $P$ (parity) nor $T$ (time-reversal) symmetry, but having the composed $PT$ symmetry.\cite{jankowski_quantizedshift} This represents a quantized shift invariant for the transient photocurrent induced by circularly-polarized light, which differs from our analysis of the linear photogalvanic effect.

 \begin{acknowledgments}

My heartfelt gratitude goes to Chong Wang who acted graciously as my sounding board, and to Boris Sturman for patiently fielding endless questions on research he accomplished four decades ago. This work has benefitted from discussions with Penghao Zhu, Takahiro Morimoto, Liang Tan, Joel Moore, Gao Lingyuan, Jay Sau, Ahn Junyeong, Joshua Deutsch and Aleksandra Nelson.

 \end{acknowledgments}

\appendix

\section{Topological invariants that depend on the inter-band Berry connection}\la{app:invariants}

This appendix section answers three related questions: (i) 
What exactly is meant by `topological invariance' if the invariant depends on the inter-band Berry connection? (ii) What exactly is meant by `continuously tuning' in Proposition (P1)? What are the conditions that preclude a discontinuous change in  $Vort_x{+}2 {RTP}_v$ in the shift-obstruction relation [\q{shiftobstruction}]?\\

In the common use of `continuously tuning',  continuity (with respect to $\bk$) is imposed on the intra-band Berry connection of the valence band, and guaranteed by the assumption that the band gap $E_g(\bk)$ is nonzero. If a nonvanishing gap (throughout the BZ or some cross-section of it) is a sufficient condition for an invariant to be insensitive to symmetric Hamiltonian perturbations, such  invariant (e.g., $RTP_v$) will be called a \emph{intra-band invariant}.  \\

In optical phenomenon, we encounter non-intra-band invariants such as $Vort_x$ whose definition assumes not only that the wave function is continuous over $BZ/2$ (as guaranteed by a nonzero band gap), but also that the inter-band Berry connection $A_{xcv\bk}{=}\braket{u_c}{i\partial_{k_x}u_v}_{cell}$ is continuous at all mirror-invariant $\bk$.   It is possible for $A_{xcv}$ to diverge when the band gap goes to zero, as is evident from the identity \q{Acvidentity}.
 Even if the band gap were everywhere nonzero,  the existence of optical vortices would make $A_{xcv}$ vanishing and discontinuous.  Both types of discontinuities are ruled out at a $\bk$-point if both the energy gap and optical affinity are nonvanishing at that $\bk$-point. \\

This discussion motivates a new definition: if the nonvanishing of the gap (in some BZ region) and the nonvanishing of the affinity (in a possibly distinct BZ region) are sufficient conditions for  an invariant to be insensitive to perturbations, such invariant (e.g., $Vort_x,\scrs_{0}$) that is not an intra-band invariant  will be called an \emph{inter-band invariant}. $Vort_x$ relies on the gap being nonvanishing over $BZ/2$ and the affinity being nonvanishing for all mirror-invariant $\bk$, while $\scrs_{0}$ relies on both gap and affinity being nonvanishing for wavevectors with $k_x{=}0$.\footnote{This difference is because $\scrs_{0}$ is an integral of the shift vector which is gauge-invariant (hence uniquely-defined) at each $\bk$; on the other hand, $Vort_x$ is an integral of $\partial_{k_y}\arg A_{xcv}$ which is not gauge-invariant at each $\bk$; to uniquely define $Vort_x$ requires that both valence-band and conduction-band wave functions be analytic over $BZ/2$.} Because the shift obstruction  relies on the insensitivity of the relative intercellular shift, the obstruction is also an inter-band invariant.\\

\begin{figure}[h]
\centering
\includegraphics[width=6 cm]{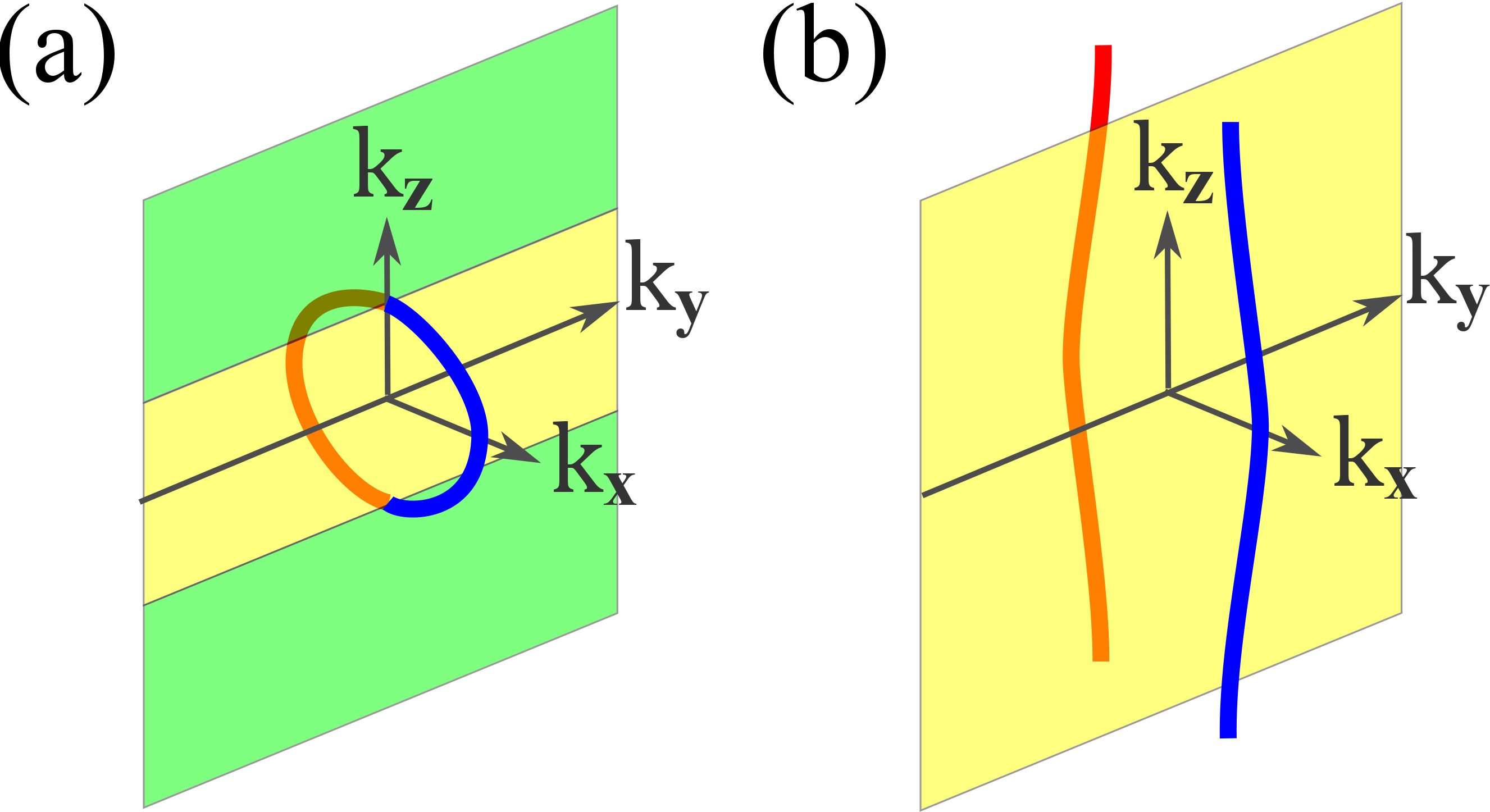}
\caption{(a) Intersection of a loop of optical zeros with the mirror-invariant $\bk$-plane at $k_x=0$. Red and blue distinguish between different segments of the optical-zero loop with opposite circulations. The net vorticity $Vort_x(k_z)$ is discontinuous at two values of $k_z$ where the yellow pane meets either green pane. (b) Representative example of optical-zero loops that extend across a nontrivial cycle of the Brillouin torus.}\label{fig:vortices3DBZ}
\end{figure}

 For three-dimensional, essentially noncentric insulators with mirror-invariant $\bk$-planes, all inter-band invariants ($Vort_x,\Delta \scrs$) in the shift-obstruction relation are generally piecewise-continuous, integer-valued functions of a third wavenumber $k_z$.
 Discontinuities can occur at  isolated values of $k_z$ where a line of optical zeros intersects the mirror-invariant $\bk$-plane, as illustrated in \fig{fig:vortices3DBZ}(a); the intersection point may be viewed as the merging of  two optical vortices with opposite circulation, as distinguished by red and blue in \fig{fig:vortices3DBZ}(a).
 The $k_z$ dependence of inter-band invariants can be ignored for the 3D insulating models explored in \s{sec:models}, which are all made from stacking 2D insulators in the $z$ direction with weak inter-layer coupling; a representative example is illustrated in \fig{fig:vortices3DBZ}(b).  For essentially noncentric (semi)metals, the intra-band invariant $RTP_v$ may also be a piecewise-continuous, integer-valued function of $k_z$.
 
\section{Greater variety of essentially noncentric insulators}\la{sec:extensions}

The main principles of essentially noncentric insulators have been formulated and exemplified in the simplest context, which however involves a few restrictive assumptions:  (i) a Bravais lattice with a monatomic basis, (ii) a reduced Hilbert space of two bands, (iii) a point group generated by a single reflection. The first restriction is relaxed in \s{sec:beyondmonatomic} and the last two in \s{sec:beyond2band}. It is hoped that a greater variety of essentially noncentric insulators increases the eventual probability of finding a material realization. 

\subsection{Beyond a monatomic basis}\la{sec:beyondmonatomic}

Thus far, I have assumed that the reduced Hilbert space is spanned by two Wannier orbitals per unit cell, and that the two orbitals in one representative unit cell are centered on the same location/site.  (This restriction need not apply to Wannier orbitals outside the reduced Hilbert space.) Here we relax the spatial restriction and allow the two orbitals (in one representative cell) to be centered at different locations, subject to the constraints imposed by the space group.\\

 Suppose the reflection-even orbital $\varphi_e$ is centered at position $\bw_e$, and $\varphi_o$ at $\bw_o$, then the tight-binding Hamiltonian becomes nonperiodic in translations by reciprocal lattice vector: $h(\bk{+}\bG){=} e^{-i\bG\cdot \bw}h(\bk)e^{i\bG\cdot \bw}$. $\bw$ here is a diagonal matrix with diagonal elements $\bw_e$ and $\bw_o$. This nonperiodic relation is necessary\cite{nelsonAA_delicatetopology}  to maintain the mod-one equivalence between the tight-binding-approximated Berry-Zak phase and the Wannier center [cf.\ \q{zakwannier}], and justifies our interpretation of the tight-binding-approximated shift vector as a positional displacement.   \\

Generically, the line-averaged shift [\q{averageshift}] at a reflection-invariant value for $k_x$ is no longer integer-valued: $\scrs_{0}{=}_1 (y[\varphi_{p(c,0)}]{-}y[\varphi_{p(v,0)}])/R_y$. (I remind the reader that $y[\varphi_e]$ is the $y$-component of the Wannier-center position $\bw_e$, and $p(b,0)$ is the parity of the Bloch state in band $b$ and with wavenumber $k_x{=}0$.) However, if the valence-band parities of both mirror-invariant $\bk$-lines are identical: $p(v,0){=}p(v,\pi/R_x)$, and likewise for the conduction band: $p(c,0){=}p(c,\pi/R_x){\neq}p(v,0)$, then 
 the shift obstruction relation [\q{shiftobstruction}] holds, with $RTP_v$ and $\Delta \scrs$ remaining integer-valued; this follows from  a simple generalization of the proof in \s{sec:theory}. The case of identical valence-band parities $[p(v,0){=}p(v,\pi/R_x)]$ is exemplified by the first-class model in \s{sec:modelfirstclass}, implying that the previous assumption of a monatomic basis is not needed for the quantization of $\Delta \scrs$.  \\

It would seem for models with identical valence-band parities that proposition (P1) is preserved but (P2) lost. However, a statement exists for models (with or without identical parities) that is a close analog of (P2):\\

\noi{P2'} For essentially noncentric, 2D insulators with a reflection symmetry, a geometric quantity exists that inputs band wave functions over a reflection-invariant $\bk$-line (say, $k_x{=}0$)  and outputs an integer $\breve{\scrs}_0$ with the following meaning:  when a mirror-invariant  quasiparticle (with $k_x{=}0$) is  optically excited, it is displaced (on average) by $\scrs_{0}R_y\vec{\by}$ in the direction of the polar axis, with  $\scrs_{0}$  that is generically non-integer-valued. This displacement vector connects the center of a reflection-even Wannier orbital $\varphi_e'$ with the center of a reflection-odd Wannier orbital $\varphi_o'$. In the standard tight-binding formalism, each Wannier orbital $\varphi$ in the tight-binding Hilbert space is assigned to a primitive unit  cell centered at a  Bravais lattice vector $(\,n_x[\varphi]R_x,\,n_y[\varphi]R_y\,)$, with $n_{x}$ and $n_y{\in}\Z$. $\breve{\scrs}_0{=}n_y[\varphi_e']{-}n_y[\varphi_o']$ if the conduction-band parity $p(c,0)$ is even; otherwise, $\breve{\scrs}_0{=}n_y[\varphi_o']{-}n_y[\varphi_e']$.\\

\noindent I refer to  $\breve{\scrs}_0$ as the \emph{generalized intercellular shift}. To define $\breve{\scrs}_0$ in terms of the band wave function: suppose two representative orbitals $\varphi_e$ and $\varphi_o$ with respective positions $\bw_e$ and $\bw_o$ are assigned to the same cell, i.e., $n_j[\varphi_e]{=}n_j[\varphi_o]$ for $j{=}x$ and $y$. Then perform a unitary transformation on the $\bk$-nonperiodic Hamiltonian so as to translate $\varphi_o$ to lie atop $\varphi_e$: $h(\bk){\ri} \breve{h}(\bk){=} U(\bk)^{-1}h(\bk)U(\bk)$, with $U(\bk)$ a diagonal matrix with diagonal elements $1$ and $e^{i\bk\cdot(\bw_e-\bw_o)}$.  $\breve{h}(\bk)$ is a $\bk$-periodic Hamiltonian with the same band energies as $h(\bk)$, but with a modified wave function denoted by $\ket{\breve{u}_{c\bk}}$ and $\ket{\breve{u}_{v\bk}}$. Then $\breve{\scrs}_0$ is defined exactly as ${\scrs}_0$ in \q{averageshift}, but with the functional dependence on $u_{b\bk}$ replaced by a functional dependence on $\breve{u}_{b\bk}$.\footnote{A generalized reverting Thouless pump invariant can also be similarly defined with $u_{b\bk}$ in \q{defineRTP} replaced by $\breve{u}_{b\bk}$. This generalization extends the meaning of a reverting pump beyond what has been considered in previous literature.\cite{AA_teleportation,nelsonAA_multicellularity,nelsonAA_delicatetopology} In the language developed in \ocite{nelsonAA_delicatetopology}, the generalized pump exists assuming the    the `mutually disjoint' symmetry condition, but not needing the `iso-orbital' condition. A physical implication of the generalized pump is the existence of surface states that interpolate across the bulk gap, for an ideal (non-relaxed, non-reconstructed) surface termination that is compatible with the chosen unit cell.\cite{nelsonAA_delicatetopology}  } In the particular case that $\bw_e{=}\bw_o$, the unitary matrix is trivial, and the generalized intercellular shift reduces exactly to the previously-defined intercellular shift in \q{averageshift}.

\subsection{Beyond two-band, reflection-symmetric Hamiltonians}\la{sec:beyond2band}

For the purpose of counting,  one band corresponds to a linearly-independent Bloch function over the BZ. For an $(N{>}2)$-band Hamiltonian with $N_c$ conduction bands (indexed by $c_1,\ldots,c_{N_v}$) and $({N_v}{=}N{-}N_c)$ valence bands (indexed by $v_1,\ldots,v_{N_v}$), I define the \emph{$N$-band intercellular shift} by summing over all inter-band intercellular shifts between the valence and conduction subspaces: 
\e{\scrs^{(N)}_{0}{=}\sum_{i=1}^{N_c}\sum_{j=1}^{N_v}\int  S^x_{yc_iv_j(0,k_y)}\f{dk_y}{2\pi}.\la{defineNbandshift}}
The utility of this definition is that if all Bloch states (in the conduction subspace, and with wavenumber $k_x{=}0$) are parity-even, and all Bloch states  (in the valence  subspace, and  with $k_x{=}0$) are parity-odd, then the $N$-band intercellular shift remains quantized to integer values.\footnote{Proof of quantization: from the definition of the shift vector in \q{defineshiftvector} and the definition of the Berry-Zak phase in \q{definezakphase}, we obtain \e{2\pi\scrs^{(N)}_{0}{=}{-}\sum_{ij}\partial_{k_y}\arg A_{xc_iv_j(0,k_y)} dk_y{+}{N_v}\sum_iZ_{c_i,0}{-}N_c\sum_jZ_{v_j,0}.} The first of the three terms is a sum of phase winding numbers and therefore takes values in $2\pi \Z$. Under the just-stated assumption on the parities of the Bloch states, $\sum_{i=1}^{N_c}Z_{c_i,0}/2\pi{=}_1\sum_{i=1}^{N_c} y[\varphi^i_e]/R_y$, with $\{\varphi^i_e\}_{i=1}^{N_c}$ labelling all reflection-even Wannier orbitals in a representative primitive unit cell. (If this identity is not apparent to the reader, I recommend Sec. VIII-C in \ocite{nelsonAA_delicatetopology} for a closely analogous proof with greater detail.) Likewise, $\sum_{j=1}^{{N_v}}Z_{v_j,0}/2\pi{=}_1\sum_{j=1}^{N_v} y[\varphi^j_o]/R_y$ for the reflection-odd Wannier orbitals in the same representative unit cell. For a monatomic basis of the Bravais lattice, $y[\varphi^i_e]{=}y[\varphi^j_o]$ for all $i$ and $j$, hence ${N_v}\sum_{i=1}^{N_c}Z_{c_i,0}{-}{N_c}\sum_{j=1}^{N_v}Z_{v_j,0}{=}2\pi \Z$, completing the proof for $\scrs^{(N)}_{0}{\in}\Z$. If the assumption of a monatomic basis is relaxed, the generalized $N$-band intercellular shift $\breve{S}_N(0){\in}\Z$ has the meaning of the net change in the primitive unit cell label when all ${N_c}{N_v}$ inter-band optical excitations are accounted for.} (This statement holds as well if `odd' is interchanged with `even'.) If the just-mentioned parity condition applies also to Bloch states with $k_x{=}\pi/R_x$, then there exists an $N$-band shift obstruction relation:\footnote{To derive this relation, apply Stoke's theorem to convert line integrals of the intra-band Berry connection $A_{ybb}$ to area integrals of the intra-band Berry curvature $\Omega_{zb}$. Then apply the complementary relation between the curvatures of conduction and valence bands: $\sum_{j=1}^{N_v}\Omega_{zv_j\bk}{=}{-}\sum_{i=1}^{N_c}\Omega_{zc_i\bk}$,\cite{spinorbitfree_AAchen} which leads to $RTP_v{=}{-}RTP_c$.}
\e{\scrs^{(N)}_{0}{-}\scrs^{(N)}_{\pi/R_x}= \sum_{i=1}^{N_c}\sum_{j=1}^{N_v}Vort_{xc_iv_j} + N\,RTP_v \in \Z. \la{Nbandshiftobstruction}}
$\scrs^{(N)}_{\pi/R_x}$ is defined as in \q{defineNbandshift} but with $0$ replaced by $\pi/R_x$; $Vort_{xb'b}=$
\e{\int \bigg(\partial_{k_y}\arg A_{xb'b({\pi}/{R_x},k_y})- \partial_{k_y}\arg A_{xb'b(0,k_y)}\bigg)\f{dk_y}{2\pi} \in \Z}
 is the net optical vorticity between bands $b'$ and $b$, and 
\e{RTP_v=\sum_{j=1}^{N_v}\int_{BZ/2}\Omega_{zv_j} \f{d^2k}{2\pi} \in \Z} 
is the returning Thouless pump of the $N_v$-band valence subspace. The reader may verify that \q{Nbandshiftobstruction}  reduces to the previously-obtained shift obstruction relation [\q{shiftobstruction}] for $N{=}2$.\\  

To recapitulate from a broader perspective, we began with a topological invariant that was previously defined for an $M$-band Hamiltonian, and were \emph{conditionally} able to extend the meaning of this invariant to an $(N{>}M)$-band Hamiltonian. This condition specifies the allowable symmetry representations for all $N$ bands in both conduction and valence subspaces. Conversely stated, the condition may be violated by adding a (topologically trivial) band with a disallowed symmetry representation to either conduction or valence subspace.  The consequence of violating the condition is that the $N$-band intercellular shift is no longer quantized to integer values. These, in a nutshell, are the hallmark attributes of    \emph{symmetry-protected delicate topology} -- a notion that has been studied for intra-band  invariants\cite{nelsonAA_multicellularity,nelsonAA_delicatetopology} but is hereby extended to inter-band invariants.\\

Currently all known examples of delicate topological insulators are essentially noncentric,\footnote{Beyond insulators, there exists a phononic three-band touching point that is both  delicate-topological and compatible with centrosymmetry.\cite{park_acoustictriplepoint} It may be possible to generalize the homotopy invariant of this three-band touching to a centrosymmetric, three-band insulator.} in the sense that the topological distinction between trivial and nontrivial insulators (as distinguished by intra-band invariants) is only meaningful for space groups without centrosymmetry.\cite{Hopfinsulator_Moore,AA_teleportation,nelsonAA_multicellularity,nelsonAA_delicatetopology,lapierre_nbandhopf}  This offers a rich playing field to search for  inter-band invariants related to the shift current. To give a flavor of the possibilities, the reverting Thouless pump (RTP) has been theoretically explored in a wide  variety of $Pn$-symmetric Hamiltonians,\cite{AA_teleportation,nelsonAA_multicellularity,nelsonAA_delicatetopology} where an $n$-fold rotational symmetry plays a role  analogous to the reflection symmetry in this paper. A known mod-$n$ equivalence\cite{nelsonAA_multicellularity,nelsonAA_delicatetopology} between the RTP and Hopf invariants suggests the existence of mod-$2n$ shift obstruction relations that relate the intercellular shift, the Hopf invariant and the optical vorticity. In this context, the intercellular shift is defined by averaging the shift vector over rotation-invariant $\bk$-lines, rather than a mirror-invariant cross-section of the BZ [cf.\ \q{averageshift}].

\section{The tight-binding approximation of the shift current: justification and pitfalls}\la{sec:TBapprox}

Having alluded to subtleties/dangers of the tight-binding approximation of shift quantities, I now elaborate on the nature of this approximation [\s{sec:nature}], provide a semi-empirical [\s{sec:semiempirical}] and rigorous [\s{sec:rigorousjustify}] justification for the approximation, and finally highlight an under-appreciated pitfall of the approximation that is specific to two-band tight-binding models [\s{sec:pitfall}].

\subsection{Nature of the approximation}\la{sec:nature}

The shift connection is expressible in terms of the matrix elements of the non-Abelian Berry connection. In the rigorously justified theory involving a Schr\"odinger-type Hamiltonian,\cite{sipe_secondorderoptical} the Berry connection is defined by
 $\tilde{\bA}_{ll'}{=}\braket{\tilde{u}_{l\bk}}{i\nabk \tilde{u}_{l'\bk}}_{cell}$, with $\tilde{u}_{l\bk}(\br){=}\tilde{u}_{l\bk}(\br{+}\bR)$ the intracell component of the Bloch function that is periodic in lattice translations, and $\br$ a \emph{continuous} spatial coordinate within the primitive unit cell. However, throughout this work, I have
approximated the Berry connection as ${\bA}_{ll'}{=}\braket{{u}_{l\bk}}{i\nabk {u}_{l'\bk}}_{cell}$, with ${u}_{l\bk}(\alpha)$ the eigenvector of an $N$-band tight-binding Hamiltonian, and $\alpha$ a \emph{discrete} intracell coordinate taking only $N$ values.  The error $\tilde{\bA}{-}\bA$ in the \emph{discrete-space approximation}  has an explicit expression [Eq.\ (B8) in \ocite{AA_wilsonloopinversion}] in terms of matrix elements of the continuous-position operator in the basis of Wannier orbitals (there being $N$ such orbitals per primitive unit cell); the approximation is equivalent to dropping all off-diagonal elements of the position operator in the just-mentioned Wannier basis -- a point of view emphasized in \ocite{julen_wannierinterpolation}. Because $\tilde{\bA}{-}\bA$ requires a correction, there is an analogous correction (derived explicitly in \ocite{julen_wannierinterpolation}) to approximating the photonic shift connection $\tilde{C}^j_{ib'b}{=}|\tilde{A}_{jb'b}|^2 \tilde{S}^j_{ib'b}$ by ${C}^j_{ib'b}{=}|{A}_{jb'b}|^2 {S}^{j}_{ib'b}$; here and henceforth, $\tilde{O}$ is defined by $O[\tilde{u}_{l\bk}]$, for $O$ that was previously defined  as a functional of $u_{l\bk}$.

\subsection{Semi-empirical justification of the approximation}\la{sec:semiempirical}

The {discrete-space approximation} is generally uncontrolled, in the sense that no known small parameter exists to bound the error: $\delta C{=}\tilde{C}^j_{ib'b}{-}{C}^j_{ib'b}$. (A small parameter exists in specific cases, as elaborated in the next \s{sec:rigorousjustify}.) The next-best course of action is to compare $\delta C$ to $\tilde{C}$  in ab-initio-based studies where Wannier functions of a continuous spatial coordinate can be accurately obtained. These studies have been carried out for a number of materials;\cite{julen_wannierinterpolation,julen_assessingrole} the most severe relative error [in the discrete-space approximation of $\tilde{\sigma}^{j}_{i\omega}$] is reported as ${\approx}50 \%$ for BC$_2$N, and for frequencies close to a band-edge excitation;\cite{julen_assessingrole} the error is significantly milder over most other frequencies, and this holds for the other material case studies as well.    
A plausible conclusion from these studies is that it is safer for a tight-binding theorist to report a value of  $\int \sigma^j_{i\omega}d\omega$ (integrated over a frequency range comparable to the bandwidth) rather than $\sigma^j_{i\omega}$ at specific frequencies -- this being another motivation for my choice of the figure of merit $F^{j}_i$ in \q{bandshift}. This point of view is not universally adopted.\cite{cook_designprinciples}\\ 

\subsection{Rigorous justification of the approximation}\la{sec:rigorousjustify}

There is at least one context where $\delta C$ is demonstrably negligible relative to $\tilde{C}$ -- in the proximity to a first-class phase transition in essentially noncentric insulators [\s{sec:modelfirstclass}]. More precisely, there exists a small parameter $s$ (proportional to the square root of the minimal energy gap $E_g$) that allows to asymptotically compare $\delta C$ and ${C}$; one can prove that  $\delta C/{C}{\sim} s$ as $s{\ri} 0$. This implies not only that the asymptotic behavior 
$F_y^x{\sim}(E_g)^{-1/2}$ [proposition (Q2)] is preserved if $\delta C$ is accounted for, but also that the coefficient $c_1$ in $F_y^x{\approx}c_1(2/E_g)^{1/2}$ is unchanged by $\delta C$. 
The conclusion that $\delta C$ is asymptotically irrelevant possibly generalizes to more classes of topological phase transitions, since the limit of vanishing energy gap is also the limit of long spatial wavelength, rendering   short-wavelength variations [of $u_{l\bk}(\br)$ within a unit cell] asymptotically irrelevant.  

\begin{widetext}

\subsection{The discrete-space approximation for two bands}\la{sec:pitfall}

The discrete-space approximation of the shift conductivity  is especially dangerous when used in conjunction with \emph{two}-band,\footnote{To clarify, time-reversal symmetry imposes that the minimal tight-binding model of an insulator has four bands, counting spin. I assume that the spin-orbit interaction is negligible and focus on one spin sector having only two bands. The Kraut-von Baltz selection rule (explained below) applies for negligible spin-orbit interaction.} time-reversal-invariant tight-binding models.  Even if resonant excitations occur only between two bands, the shift connection generally receives contributions from virtual excitations to other intermediate bands, as has been made explicit by sum-over-states formulas in \ocite{baltz_bulkPV} and \ocite{cook_designprinciples}:
\e{\tilde{C}^j_{ib'b\bk}\eq -\text{Im} \f{\overline{\tilde{v}^j_{b'b}}}{(\var_{b'b})^2}\bigg[\braopket{\tilde{u}_{b'}}{\partial_{k_i}\partial_{k_j}H}{\tilde{u}_b}_{cell}-\f{\tilde{v}^i_{b'b}\Delta^j_{b'b}+\tilde{v}^j_{b'b}\Delta^i_{b'b}}{\var_{b'b}}+\sum_{{b''}\neq {b'},b}\bigg(\f{\tilde{v}^i_{{b'}{b''}}\tilde{v}^j_{{b''}b}}{\var_{{b'}{b''}}}-\f{\tilde{v}^j_{{b'}{b''}}\tilde{v}^i_{{b''}b}}{\var_{{b''}b}}\bigg) \bigg],\la{sumover}}
with $\tilde{v}^j_{b'b\bk}{=}i\braopket{\tilde{u}_{b'}}{\emikr [\hat{H},r^j] \eikr}{\tilde{u}_b}_{cell}/\hbar $ being a matrix element of the $b$'th component of the velocity operator, $\var_{b'b}{=}\var_{b'\bk}{-}\var_{b\bk}$ being a difference in band energies,  $H(\bk){=}\emikr \hat{H} \eikr$ being the single-particle Bloch Hamiltonian, and $\Delta^i_{b'b}{=}\partial_{k_i} \var_{b'}{-}\partial_{k_i}\var_n$ being a difference in band velocities. \\

I assume that the Schr\"odinger-type Hamiltonian has the form $\hat{H}{=}p^2/2m{+}U(\br,\bp)$ with $U$ that is at most linear in the canonical momentum $\bp$.\footnote{As was emphasized in \ocite{chong_firstprinciplesnonlinear}, $\braopket{\tilde{u}_{b'}}{\partial_{k_i}\partial_{k_j}H}{\tilde{u}_b}_{cell}$ might be nonzero  for ab-initio calculations where the pseudopotential $U$ may be nonlinear in $\bp$. This is in principle one way to evade the selection rule, though further quantitative studies are needed to quantify this evasion.}
Under this assumption, $\braopket{\tilde{u}_{b'}}{\partial_{k_i}\partial_{k_j}H}{\tilde{u}_b}_{cell}$ vanishes for $b'{\neq} b$, and the \emph{longitudinal shift conductivity} ($\tilde{\sigma}^{i}_i$) vanishes if one neglects all `virtual excitations', i.e.,  if one neglects any excitation to an intermediate band that is not either of the two bands of greater interest. Precisely, I mean that the summation term on the right-hand side of \q{sumover} is much smaller than the middle term on the right-hand side; this assumption may hold  when the band-energy difference $|\var_{b'b}|$ is much smaller than $|\var_{b'b''}|$ and $|\var_{bb''}|$, for any $b''\neq b,b'$.  The vanishing of $\tilde{\sigma}^{i}_i$ under these assumptions was first proven generally in \ocite{baltz_bulkPV} and will thus be called the \emph{Kraut-von Baltz selection rule}. A more precise statement is that $\tilde{C}^{i}_{ib'b\bk}{=}0$ if virtual transitions are ignorable for that particular value of $\bk$, as can be verified from \q{sumover} if the first and third terms (on the right-hand side) are dropped. In ab-initio-derived models, it is possible that $\tilde{C}^{i}_{ib'b\bk}$ approximately vanishes over some regions of the Brillouin zone where $|\var_{b'b\bk}|$ become unusually small, while remaining nonzero in other regions. \\

Unfortunately, the selection rule has been under-appreciated\cite{LiangTan_upperlimit} or mis-interpreted\cite{cook_designprinciples} in recent works that purport to predict a value  for $\sigma^{i}_i$ (or upper limit for $\int \sigma^{i}_{i\omega}d\omega$) based on two-band tight-binding models. In interpreting either of these works, one can take one of two positions: \\

\noi{i} Suppose virtual excitations are exactly zero, then $\tilde{\sigma}^{i}_i{=}0$, according to Kraut-von Baltz. It is also possible for the two-band tight-binding approximation (${\sigma}^{i}_i$) to be nonzero. Indeed, the two-band-tight-binding-approximated shift connection is expressible in a form closely analogous to \q{sumover}:
\e{{C}^i_{ib'b\bk}\eq -\text{Im} \f{\overline{v^i_{b'b}}}{(\var_{b'b})^2}\bigg[\braopket{{u}_{b'}}{\partial_{k_i}^2h_{2-band}}{{u}_b}_{cell}-\f{v^i_{b'b}\Delta^i_{b'b}+v^i_{b'b}\Delta^i_{b'b}}{\var_{b'b}} \bigg]\la{sumover2}} 
with $h_{2-band}(\bk)$ a two-band tight-binding Hamiltonian;
one finds that $C^{i}_i$ can be nonzero  because $\braopket{{u}_c}{\partial^2_{k_i}h_{2-band}}{{u}_v}_{cell}$ is
generically nonzero, as was correctly argued in \ocite{cook_designprinciples}. $C^{i}_i{\neq}0$ and $\tilde{C}^{i}_i{=}0$ are manifestly consistent statements, implying that the correction  $\delta C{=}\tilde{C}_{ib'b}^{i}{-}{C}_{ib'b}^{i}$ (explicitly derived by Ibanez-Azpiroz-Tsirkin-Souza\cite{julen_wannierinterpolation}) exactly cancels ${C}_{ib'b}^{i}$. This potentially surprising cancellation follows from adopting a pathological assumption. \\

\noi{ii} Suppose virtual excitations to a third band are nonzero, then the Kraut-von Baltz selection rule does not hold. Then any two-band,  tight-binding Hamiltonian cannot be a complete model of the longitudinal shift current, and any expression\cite{cook_designprinciples} (or upper limit\cite{LiangTan_upperlimit}) for $\sigma^{i}_i$ that depends only on parameters of a two-band, tight-binding Hamiltonian has questionable value.\footnote{Naively equating \q{sumover} with \q{sumover2}, one may be tempted to interpret  the $\braopket{{u}_{b'}}{\partial_{k_i}^2h_{2-band}}{{u}_b}_{cell}$ term as encoding the virtual transitions outside of the two-band subspace:
\e{ \braopket{{u}_{b'}}{\partial_{k_i}^2h_{2-band}}{{u}_b}_{cell}  \as \substack{?\\ \approx}\as  \sum_{{b''}\neq {b'},b}\bigg(\f{\tilde{v}^i_{{b'}{b''}}\tilde{v}^i_{{b''}b}}{\var_{{b'}{b''}}}-\f{\tilde{v}^i_{{b'}{b''}}\tilde{v}^i_{{b''}b}}{\var_{{b''}b}}\bigg). } 
However, the exact formula for a general $N$-band tight-binding Hamiltonian is 
\e{{b'}\neq b:\as \braopket{{u}_{b'}}{\partial_{k_i}^2h_{N-band}}{{u}_b}_{cell}= \var_{b'b}(i\partial_{k_i}+A_{i{b'}{b'}}-A_{ibb}) A_{ib'b}\lin
 +\f{v^i_{b'b}\Delta^i_{b'b}+v^i_{b'b}\Delta^i_{b'b}}{\var_{b'b}}-\sum_{{b''}\neq {b'},b}\bigg(\f{v^i_{{b'}{b''}}v^i_{{b''}b}}{\var_{{b'}{b''}}}-\f{v^i_{{b'}{b''}}v^i_{{b''}b}}{\var_{{b''}b}}\bigg),\la{sumover3}} 
which differs substantially from the hypothesized interpretation. Moreover, for a two-band Hamiltonian, the summation term in \q{sumover3} drops out, expressing the simple fact that one needs a $(N{>}2)$-band Hamiltonian to describe virtual transitions outside of a two-band subspace.}\\


 It is worth remarking that even a minor absolute error in mis-calculating the longitudinal shift connection  is amplified to infinity  in a calculation of the longitudinal shift conductivity, if the joint density of states diverges -- which unfortunately was the case in the band-edge calculations of \ocite{cook_designprinciples}.  \\

Thus if a two-band tight-binding model is the preferred method, it is safer to predict the \emph{transverse shift conductivity} $\int \sigma^{j}_{i\omega}d\omega$ (with $i{\neq}j$) rather than the longitudinal conductivity $\int \sigma^{i}_{i\omega}d\omega$. This is one motivation for why only the transverse conductivity $\int\sigma^{x}_{y\omega}d\omega$ (with $y$ parallel to the polar axis) was explicitly calculated for the two-band models in \s{sec:models}. 

\end{widetext}

\bibliography{bib_Apr2018}

\end{document}